\documentclass[12pt]{article}
\usepackage{amsmath,amssymb,latexsym}
\usepackage[mathscr]{eucal}
\usepackage[dvips]{graphicx}
\usepackage{feynmf,subfigure}

\newcommand{\Z}{{\mathbb Z}}

\newcommand{\beq}{\begin{equation}}
\newcommand{\eeq}{\end{equation}}
\newcommand{\beqa}{\begin{eqnarray}}
\newcommand{\eeqa}{\end{eqnarray}}

\newcommand{\uno}{\mbox{1\kern-.59em {\rm l}}}

\newcommand{\be}{\begin{equation}}
\newcommand{\ee}{\end{equation}}
\newcommand{\bea}{\begin{eqnarray}}
\newcommand{\eea}{\end{eqnarray}}

\begin{document}
\begin{fmffile}{diagB}

 \fmfcmd{%
     vardef cross_bar (expr p, len, ang) =
      ((-len/2,0)--(len/2,0))
         rotated (ang + angle direction length(p)/2 of p)
         shifted point length(p)/2 of p
     enddef;
 vardef doublecross_bar_third (expr p, len, ang) =
     ((-len/2,0)--(len/2,0))
         rotated (ang + angle direction length(p)/3 of p)
         shifted point length(p)/3 of p
             enddef;
 vardef doublecross_bar_twothird (expr p, len, ang) =
    ((-len/2,0)--(len/2,0))
         rotated (ang + angle direction 2length(p)/3 of p)
         shifted point 2length(p)/3 of p
     enddef;
     style_def cross_gauge expr p =
       cdraw (wiggly p);
       ccutdraw cross_bar (p, 5mm,  45);
       ccutdraw cross_bar (p, 5mm, -45)
     enddef;
     style_def cross_scalar expr p =
       draw_dashes p;
       ccutdraw cross_bar (p, 5mm,  45);
       ccutdraw cross_bar (p, 5mm, -45)
      enddef;
      style_def doublecross_gaugethird expr p =
       cdraw p;
      ccutdraw doublecross_bar_third (p, 5mm,  45);
       ccutdraw doublecross_bar_third (p, 5mm, -45)
     enddef;
     style_def doublecross_gaugetwothird expr p =
       cdraw p;
       ccutdraw doublecross_bar_twothird (p, 5mm, 45);
       ccutdraw doublecross_bar_twothird (p, 5mm, -45)
     enddef;
}

\begin{titlepage}
\begin{flushright}
{ROM2F/2006/07}\\
\end{flushright}
\begin{center}
\vspace{2.2cm}
{\large \sc Extra-Dimensions and Dark Matter}\\
\vspace{2.2cm}
{\sc Francesco Fucito, Andrea Lionetto and Maurizio Prisco}\\
\vspace{1cm}
{\sl Dipartimento di Fisica, Universit\`a di Roma ``Tor Vergata''\\
I.N.F.N. Sezione di Roma II,\\
Via della Ricerca Scientifica, 00133 Roma, Italy}\\
\end{center}
\vskip 1.5cm
\begin{center}
{\large \bf Abstract}
\end{center}
{In this paper we study the general scenario of an effective theory coming from the
compactification of a higher dimensional theory in a string inspired setting.
This leads to gauge coupling unification at an intermediate mass scale. After having
computed all the threshold corrections (due to Kaluza-Klein modes) to the running of the 
couplings of the MSSM we embark in a detailed phenomenological analysis of the model, based
on the numerical package DarkSUSY, to find constraints on the scenario from Dark Matter data.
The mass spectrum of the theory does not have tachyons. Moreover
we find that the neutralino is still the LSP with a relic density compatible with the most recent
experimental data. With respect to the standard mSUGRA scenario we find that the neutralino is
higgsino like in most of the parameter space.
Our modifications to the DarkSUSY package will be shortly available upon request.}

\par    \vfill
\end{titlepage}
\addtolength{\baselineskip}{0.3\baselineskip}
\setcounter{section}{0}
\section{Introduction}
\setcounter{equation}{0}
The idea of extra dimensions has brought to an explosion of activity
in the field of high energy physics phenomenology. The unification of the couplings
of the theory at an intermediate mass scale is in fact particularly appealing. Narrowing
the gap between the energy scale that can be probed by current experimental techniques and the unification 
energy scale makes indirect test of the latter more reliable. 

The connection between the supersymmetry breaking energy scale and the size of the compactified extra
dimensions first appeared in \cite{Antoniadis:1988jn,Kounnas:1988ye}. At that time the result was considered
a negative one: a breaking scale of the order of the TEV implied a compactification scale of 
$\approx 10^{18}$ GeV, a region in which perturbative computations in string theory are ruled out.
This judgement was influenced by the fact that the fashionable string theory at the time was the 
heterotic string and that the other consistent string theories were not considered to be relevant 
from the phenomenological point of view.
It was only later, with the discovery of dualities among different string theories \cite{Hull:1994ys}
that this scenario became
appealing \cite{Arkani-Hamed:1998rs}: all consistent string theories gained in fact equal status.
Non-perturbative computations for a certain string theory became perturbative ones for its dual.
Being in the perturbative regime is only one of the desirable features that the theory should have.
Another major hurdle for a model building directly coming from string theory is given by the presence 
of moduli. These are scalar fields which appear after the compactification and give flat directions to
the potential of the theory. How to fix this problem is not clear yet, even if encouraging progresses have 
been done recently \cite{KachruJR,Kallosh,Antoniadis:2004pp,Bianchi:2005yz,Lust:2005dy}.

Other inputs on the relevance of the extra-dimensions scenario for phenomenology could be given by a 
comparison with experimental data of present and sensitivities of future experiments. 
The approach in which a low energy
model is derived directly from a string theory is called top-down, while trying to prune models on the
basis of low-energy data is called bottom-up. Eventually these two approaches will merge, but for the moment
they are still distinct.

This paper is an example of the bottom-up approach: as suggested in \cite{Dienes:1998vg} a simple way to 
achieve an earlier unification is to modify the running of the couplings in an ''exponential'' fashion.
This can be accomplished by keeping into account the effects of extra massive Kaluza-Klein (KK) states. The 
scenario here is quite generic. A detailed knowledge of the compactification mechanism is not needed.
At the start only the gauge sector and the Higgs fields are allowed to have a tower of KK states.
At a later time also part of the fermionic sector can have a tower. 

 In this paper we analyse the implications of this scenario for the problem of dark matter 
using the numerical package DarkSUSY~\cite{Gondolo:2004sc} together with a modified version of the ISASUGRA~\cite{Paige:2003mg} package. In order to do this we compute the one loop running of all the couplings in the minimal SUSY extension of the standard model (MSSM) in the extra dimension scenario of Ref.\cite{Dienes:1998vg}. 
To do this efficiently we use the method of the effective 
potential \cite{Barbieri:1982nz}. See the lecture notes in Ref.~\cite{derendinger} for a nice summary.
The computation of the running of the gauge and Yukawa couplings was performed already 
in Ref.\cite{Dienes:1998vg}, while the running of the soft parameters was first 
studied in \cite{Kobayashi:1998ye} using the spurion formalism. Instead we used a diagrammatic analysis finding some discrepancies between our results and that of the previous reference. 

This is the plan of the paper: in Section~\ref{sec:extradim} we outline the theoretical framework and the method of the effective potential. In Section~\ref{sec:running} we present the renormalization group equations in presence of extra-dimensions that we implemented in the numerical package ISASUGRA. In Section~\ref{sec:pheno} we show the results of our computations: 
the spectrum of the theory has no tachyons. The neutralino remains
the LSP particle in almost all of the parameter space we have explored. If no fermions are allowed to have
a KK tower then the neutralino is higgsino-like. Otherwise it is bino-like 
for $\tan\beta< 30$ and higgsino or mixed for $\tan\beta> 30$.

\section{Extra-dimensions, renormalization and the effective potential}
\setcounter{equation}{0}
\label{sec:extradim}
In this Section we deal with the theoretical issues which must be faced to adapt the DarkSusy package
to the extra-dimensions scenario of \cite{Dienes:1998vg}. In the first subsection we review 
the main aspects of the extra-dimensions scenario of \cite{Dienes:1998vg} while in the second subsection
we review the computation of the one-loop $\beta$-functions of the MSSM with the method of the 
effective potential\cite{Barbieri:1982nz,derendinger}.
\subsection{An extra-dimensions scenario}
The MSSM emerges after the extra dimensions are compactified on circles of fixed radius $R$. 
The energy scale $R^{-1}\equiv \mu_0$ exceeds experimentally attainable energy scales and represents 
the threshold beyond which the effects of extra-dimensions can be felt. Let $\delta$ be the number
of extra-dimensions. The effect of the compactification of a $4+\delta$-dimensional field along $\delta$
circles of radius $R$ is to allow for the presence of four-dimensional massive excitations, called 
KK modes, of mass $m_n^2=m_0^2+\vec n\cdot\vec n/R^2$. $\vec n$ is a $\delta$-dimensional
vector with $n_i\in\Z$. $m_0$ is the mass of the zero-mode. If an additional $\Z_2$ is imposed on
the compactified dimensions, the circles change into orbifolds. Fields sitting at the orbifold 
fixed points are not allowed to have KK modes. This is a convenient mechanism to select which
fields have massive KK excitations. Let us now denote by $\eta$ the number of generations 
of MSSM chiral fermions. In the simplest possible scenario $\eta=0$: no chiral MSSM fermions have 
KK excitations. We will discuss scenarios with both $\eta=0$ and $\eta\neq 0$.
Compactifications with $\delta=1,2$ will exhibit $N=2$ SUSY. Higher values of $\delta$ will
result in theories with an even
larger number of supersymmetries. The $N=1$ vector multiplet, together with an additional chiral 
multiplet\footnote{This chiral multiplet must be odd under the $\Z_2$, so that its zero-modes 
are missing,
to recover the expected $N=1$ MSSM.}, $\Phi$, are such that their KK modes are organized into an 
$N=2$ multiplet. 
Finally the two chiral fields, $Q, \tilde Q$, containing the two Higgs fields form an $N=2$ hypermultiplet.
For $\eta\ge 1$ also the chiral multiplet containing one of the chiral fermions and its mirror 
are organized into an $N=2$ hypermultiplet. 
As we said earlier, this $N=2$ SUSY for the KK modes is welcome since it allows the gauge 
couplings to be computed {\it \`a la} Wilson. To make the Lagrangian of the theory really $N=2$ SUSY
invariant we must add a term, $Q\Phi\tilde Q$, which couples
the chiral field in the $N=2$ multiplet with the fields in the hypermultiplet. This term is only there
for the KK modes: zero modes have, in fact, a $N=1$ SUSY. To test how robust is our scenario to this term
we have performed computations with and without this coupling: as we will see when we will discuss our
results, there seem to be minor differences between the two cases which we will dub Higgs $N=1$ and $N=2$.
\subsection{Computations of the  one loop $\beta$-functions}
For energy scales larger than $\mu_0$, KK excitations
become possible and their effects have to be incorporated in the computation of the $\beta$-functions.
Given the large number of such functions to be computed in the MSSM we prefer to carry out this computation
using the method of the effective potential \cite{Barbieri:1982nz,derendinger}. 
In fact, in SUSY theories all divergences are found in the
wave-function renormalization of chiral multiplets and in the
renormalization of gauge coupling constants (equivalent to
wave-function renormalization of vector multiplets). 
At the one-loop level, it is then
sufficient to consider the renormalization of the scalar
potential to determine all the renormalization constants of the theory. The
one-loop divergent contributions to the scalar potential are given
by 
\be 
\label{deltaV1loop} 
\delta V=\frac{\Lambda^2}{32\pi^2}Str
{\cal M}^2(z)+\frac{1}{64\pi^2}Str {\cal M}^4(z)
\ln\left(\frac{{\cal M}^2(z)}{\Lambda^2}\right)
\ee 
where $\Lambda$ is a cut-off mass parameter. The first term contains the
quadratic divergences. $Str{\cal M}^2(z)$ is considered as a
function of the scalar fields $z^a$ and their vacuum
expectation values. $Str{\cal M}^4(z)$ is the analogous
supertrace for the fourth power of the mass matrices. The second
term contains all logarithmic one-loop divergences. 

The scalar potential is 
\begin{equation}
V(z^a,z_a)=V_{susy}(z^a,z_a)+V_{soft}(z^a,z_a) 
\label{scalarpot}
\end{equation}
where $V_{susy}$ is the standard supersymmetric potential 
\be
V_{susy}(z^a,z_a)=f_af^a+\frac{1}{2}D^A D^A 
\ee 
and where we used the standard notation $f_a=df/dz^a,  \quad f^a=df/dz_a$. 
More subscript and superscript indices denote higher order derivatives. The D-terms
read
\begin{equation}
D^A=-gz_a{T^A}^a_b z^b 
\end{equation}
in terms of the generators $T^A$ of the gauge group for the
representation of the chiral multiplets. The superpotential will be written
\begin{equation}
f=\frac{1}{2}\mu_{ab} z^az^b+\frac{1}{6}f_{abc} z^az^bz^c 
\end{equation}
Following the notation in \cite{Martin:1997ns}, we set $z^a=H_u, H_d, Q, L, \bar u, \bar d, \bar e$
for the chiral superfields of the theory.
Since the top quark, bottom quark and tau lepton are the heaviest fermions in the Standard Model,
we will use the approximation in which only the 33 component of the Yukawa matrices is important.
Then, in terms of the components with respect to the weak $SU(2)$, the relevant fields 
become $Q_3=(t, b), L_3=(\nu_\tau, \tau), H_u=(H_u^+, H_u^0), H_d=(H_d^0, H_d^-), 
\bar u_3=\bar t, \bar d_3=\bar b, \bar e_3=\bar \tau$ and the superpotential reads 
\bea
\int d^2\theta f_af^a&\approx& y_t(t\bar tH_u^0-\bar t bH_u^+)-y_b(t\bar bH_d^--\bar b bH_d^0)
-y_\tau(\nu_\tau\bar \tau H_d^--\bar \tau \tau H_d^0)\nonumber\\
&+&\mu(H_u^+H_d^--H_u^0H_d^0)
\label{supersusy}
\eea
$V_{soft}(z^a,z_a)$ keeps in account soft breaking terms. 
These are gauge invariant terms which do not add
quadratic divergences that is $Str{\cal M}^2(z)$ does not
receive any new field dependent contributions from these terms.  
Keeping in account gauge invariance we only need to add scalar and fermionic terms
\be 
V_{soft}(z^a,z_a)=(m^2)^a_b z_a
z^b+\xi(z)+\bar\xi(\bar z)  
\label{soft}
\ee 
where $\xi$ is an arbitrary gauge invariant cubic function of $z_a$
\begin{equation}
\xi(z)=\xi_{(1)a}z^a+\xi_{(2)ab}z^a z^b+\xi_{(3)abc}z^a z^b
z^c
\label{softpot}
\end{equation}
$V_{soft}(z^a,z_a)$ only adds constant terms to the squared mass matrix and
(\ref{soft})
is the most general expression for scalar soft breaking terms. They
contain scalar mass terms of the form $z\bar z$, $z^2+\bar
z^2$ and trilinear analytic interactions $z^3+\bar z^3$. The
spin $1/2$ mass matrix instead 
\be
-\frac{1}{2}\Delta^{AB}\lambda^A\lambda^B-\frac{1}{2}\bar\Delta^{AB}\bar\lambda^A\bar\lambda^B
\ee 
will only give a constant contribution
$\Delta^{AB}\Delta^{BA}$ to $tr {\cal M}_{1/2}^2$. This term is
then a soft term too. All other additions to ${\cal M}_{1/2}^2$
are not soft in general. These soft breaking terms contain all we
need to find an acceptable spectrum of SUSY particles.
We have soft terms to raise the masses of scalar quarks
and leptons, and we also have gaugino mass terms to raise the mass
of gluinos and photino.
Using the same notation as above, denoting by $a_t, a_b, a_\tau$ the 33 component of the 
matrices $\xi_{(3)uQH_u}, \xi_{(3)dQH_d}, \xi_{(3)\bar eLH_d}$ and assuming that the squark and lepton
matrices are flavour blind we get
\bea
V_{soft}
&\approx& \left[-\frac{1}{2}(\Delta^{33}\tilde g\tilde g+\Delta^{22}\tilde W\tilde W+ \Delta^{11}\tilde B
\tilde B)+c.c.\right]-m^2_{Q_3}\tilde t^*\tilde t-m^2_{Q_3}\tilde b^*\tilde b\nonumber\\
&-&m^2_{L}\tilde \nu_\tau^*\tilde \nu_\tau-m^2_{L}\tilde \tau^* \tau-m^2_u\tilde{\bar t}^* \tilde{\bar t}
-m^2_d\tilde{\bar b}^* \bar{\tilde b}-m^2_{\bar e}\tilde{\bar\tau}^* \tilde{\bar\tau}\nonumber\\
&-&a_t(t\bar tH_u^0-\bar t bH_u^+)-a_b(t\bar bH_d^--\bar b bH_d^0)
-a_\tau(\nu_\tau\bar \tau H_d^--\bar \tau \tau H_d^0)\nonumber\\
&+&\left[b(H_u^+H_d^--H_u^0H_d^0)+c.c.\right]-m^2_{H_u}\left[(H_u^+)^*H_u^++(H_u^0)^*H_u^0\right]\nonumber\\
&-&m^2_{H_d}\left[(H_u^-)^*H_u^-+(H_u^0)^*H_u^0\right]
\label{supersoft}
\eea
$\tilde g, \tilde W, \tilde B$ are the gaugino, wino and bino.
Let us now compute the
supertrace of the quartic mass matrix for arbitrary values of the
fields $z^a$. Let's start with the mass matrix for spin one
\begin{equation}
(\mathcal{M}^2_1)^{AB}=D^A_aD^B_a+D^B_aD^A_a 
\end{equation}
The fermionic mass matrix has instead the form
\bea
\mathbf{\mathcal{M}_{1/2}}= \left( \begin{array}{cc}
f_{ab} & i\sqrt 2 D^A_a \\
-i\sqrt 2 D^{Ba} & \Delta^{AB}
\end{array} \right) 
\eea
From this matrix one obtains
\begin{eqnarray}
-2 tr \mathcal{M}^4_{1/2} &=&-2f_{ab}f^{bc}f_{cd}f^{da}-16f_{ab}f^{ac}D^A_c D^{Ab}-16(D^A_aD^{Ab})(D^{Ba}D^B_b)  \nonumber \\
&&+ 8f_{ab}D^{Aa}\Delta^{AB} D^{Bb}+f^{ab}D^A_a\bar\Delta^{AB} D^B_b-16 \bar\Delta^{AC}\Delta^{CB} D^A_a D^{Ba}  \nonumber \\
&&-2\Delta^{AB}\bar\Delta^{BC}\Delta^{CD}\bar\Delta^{DA} 
\end{eqnarray}
The mass matrix squared for the scalar fields is given, as usual,
in terms of second derivatives of the potential
\bea
\mathbf{\mathcal{M}^2_0}=\left( \begin{array}{ccc}
V^a_b & V^{ac} \\
V_{db} & V^c_d
\end{array} \right) 
\eea
where
\begin{eqnarray}
V_{ab} & = & f_{abc} f^c+D^A_aD^A_b+\eta_{ab}  \\
V^{ab} & = & f^{abc} f_c+D^{Aa}D^{Ab}+\bar\eta^{ab}  \\
V^a_b & = & f^{ac} f_{bc}+{D^A}^a_bD^A+D^{Aa} D_b^A+(m^2)^a_b 
\end{eqnarray}
Note that the mass matrix squared is field independent. The soft terms give
\begin{equation}
tr \mathcal{M}^2_{soft}=2(m^2)^a_a-2\Delta^{AB}\bar\Delta^{BA} 
\end{equation}
which is also field independent. 

For the trace of the fourth power of the mass matrix, after some calculations we finally obtain
\begin{eqnarray}
Str \mathcal{M}^4 & = & +2g^2[T(R)+2C(R)-3C(G)]D^A D^A-8g^2C(R)f^af_a+ \nonumber \\
&&+4f^{ac} f_{bc} {D^A}^b_a D^A+2f^{abc} f_{abd} f^d f_c-2g^2C(R)[z^a f^b f_{ab}+z_a f_b f^{ab}]+ \nonumber \\
&&-8g^2 C(R)\Delta_A(f_a z^a+f^a z_a)-16g^2C(R)\Delta_A^2 z^a z_a-2\Delta^4_A+ \nonumber \\
&&+2f^{abc} f_c \eta_{ab}+2f_{abc} f^c \eta^{ab}-2g^2C(R)(\bar\eta^a z_a+\eta_a z^a)+2\eta_{ab} \bar\eta^{ab}+ \nonumber \\
&&+4f^{ab} f_{ac}(m^2)^c_b+4g^2C(R)(m^2)^a_b z_a z^b+2(m^2)^a_b(m^2)_a^b+ \nonumber \\
&&+4(m^2)^a_b(D^A)^b_a D^A 
\label{tracem4}
\end{eqnarray}
From (\ref{tracem4}) we can compute the coefficients of the terms with different powers of the variables $z, \bar z$.
The renormalization of the gauge coupling constant $g$ and the Yukawa
couplings $f_{abc}$ is not affected by the soft terms which do not contribute to the dimension four terms $zz\bar z\bar z$.
Then one defines the following renormalization constants
\begin{eqnarray}
\hat z^a&=&\left( \delta^a_b-\frac{1}{2}\epsilon^a_b\right)z^b \\
\hat g&=&(1+\rho)g \\
\hat f_{abc}&=&f_{abc}+\frac{1}{2}\epsilon^{a'}_a f_{a'bc}+\frac{1}{2}\epsilon^{b'}_b f_{ab'c}+
\frac{1}{2}\epsilon^{c'}_c f_{abc'}+f^{a'b'c'}_{abc}f_{a'b'c'} \\
\hat \mu_{ab}&=&\mu_{ab}+\frac{1}{2}\epsilon^{a'}_a \mu_{a'b}+
\frac{1}{2}\epsilon^{b'}_b \mu_{ab'}+\mu^{a'b'}_{ab}\mu_{a'b'}
\end{eqnarray}
with
\begin{eqnarray}
\rho=-2kg^2[T(R)-3C(G)]  \\
\epsilon^a_b=4kg^2C(R)\delta^a_b-2kX^a_b
\end{eqnarray}
where $X^a_b=f^{acd}f_{bcd}$ and
\begin{eqnarray}
\mu^{a'b'}_{ab}&=&2kg^2[C(A)+C(B)]\delta^{a'}_a\delta^{b'}_b  \\
f^{a'b'c'}_{abc}&=&2kg^2[C(A)+C(B)+C(C)]\delta^{a'}_a\delta^{b'}_b\delta^{c'}_c 
\end{eqnarray}
Furthermore
\begin{equation}
k=\frac{1}{32\pi^2} \ln (\Lambda/Q) 
\end{equation}
where $\Lambda$ is an ultraviolet cutoff and $Q$ is the energy scale at which the couplings are computed.
As it is well-known in SUSY theories the couplings renormalize as
\begin{eqnarray}
\mu_{ab}&=&({\cal Z}^{1/2})_a^{a'}({\cal Z}^{1/2})_b^{b'} \mu_{a'b'}  \\
f_{abc}&=&({\cal Z}^{1/2})_a^{a'}({\cal Z}^{1/2})_b^{b'}({\cal
Z}^{1/2})_c^{c'} f_{a'b'c'} 
\label{rencouplings}
\end{eqnarray}
From (\ref{rencouplings}) we infer
\begin{equation}
({\cal Z}^{1/2})^a_b=\delta^a_b+4kg^2C(A)\delta^a_b-kX^a_b 
\end{equation}
It is now easy to compute the renormalizations of the couplings and masses of the theory and from this
their runnings. In Appendix A we collect, for the reader's convenience, all these results 
following \cite{derendinger}. It is now easy to specify these formulae to the fields of the MSSM and
recover the results of \cite{Martin:1997ns} for the running of the MSSM couplings and masses.

In presence of KK modes the effect of the thresholds is to 
contribute to the running with power-like terms \cite{Dienes:1998vg}. The general form of the running of 
a coupling $g$ will be of the type
\begin{equation}
Q\frac{dg}{dQ}=(b-\tilde b)\ln\frac{\Lambda}{Q}+
X_{\delta}\tilde b\left(\frac{\Lambda}{\mu_0}\right)^{\delta}
\label{runnings}
\end{equation}
$X_{\delta}$ is a constant depending on the number of compactified dimensions. $b$ is the standard one loop
value of the coefficient, while $\tilde b$ is the contribution of the thresholds.

\section{Running of the couplings}
\setcounter{equation}{0}
\label{sec:running}
In this Section we list the running of the couplings of the theory. To find them we used the 
formulae of Appendix A from which we extracted the contributions of the various KK towers.
In certain cases we also checked these results against the appropriate Feynman diagrams which
renormalize the theory.
Appendix B contains some details of the computation which will be fully discussed in \cite{prisco}.

The running of the gauge and Yukawa couplings were first studied in \cite{Dienes:1998vg}: 
in this case the coefficients $\tilde{b}_a$ take into account the Higgs and gauge bosons KK tower and,
separately, the contribution of $\eta$ families of chiral fermions. The one-loop 
coefficients (\ref{runnings})
for  the running of the gauge couplings are
\bea\label{b coeff eta}
b_a&=&(\frac{33}{5},1,-3)\nonumber\\
\tilde{b}_a&=&(3/5,-3,-6)+\eta(4,4,4) 
\eea
where $a=1, 2, 3$ labels the three gauge groups $U(1), SU(2), SU(3)$. The $U(1)$ generator is rescaled
by a factor 3/5.

As we said earlier, it is convenient to assemble the two $N=1$ Higgs chiral fields into an 
$N=2$ hypermultiplet.
To be invariant under $N=2$ SUSY, besides the kinetic terms for the vector multiplet and the hypermultiplet 
the lagrangian must also contain an interaction term of the following form
\be
\int d^2\theta \left(\sqrt{2} H_u {\Phi} H_d+\mu H_uH_d\right)+h.c.
\label{eq:n=2intterms}
\ee  
where $\mu$ plays the role of the usual $\mu$-term appearing in the MSSM while $\Phi$ is the chiral 
multiplet contained in the $N=2$ vector multiplet. These are the only admissible terms for an $N=2$ 
theory. The first term is a coupling between the hypermultiplet and the vector multiplet, and it is 
needed to enlarge the amount of supersymmetry.
The one-loop diagram built from the interaction vertex associated to this term exactly cancels the 
one loop diagram of Fig.~6d of~\cite{Dienes:1998vg} (in whose loop KK states can propagate) thus 
ensuring that the Higgs wavefunction is immune to the effects of the extra space time dimensions.
The first term in (\ref{eq:n=2intterms}) is not present at the level of zero modes because  
$\Phi$ is odd under the action of the $\Z_2$ of the orbifold. In this way even if the Higgs 
zero-modes are assembled into an $N=2$ hypermultiplet they still have an anomalous dimension.

To check how robust the scenario is with respect to the presence of this term, we have studied two 
different cases which we call Higgs $N=1$ and $N=2$. In the former case we assume that the KK states 
behave as their zero-modes and that the first term in (\ref{eq:n=2intterms}) is absent.
In the latter case this term is switched on. 

\subsection{Higgs $N=1$}
\subsubsection{Couplings in the SUSY Lagrangian}
In this case the anomalous dimension receives a contribution from the diagrams in which particles from  
the KK tower of the gauge bosons circulate in the loop leading to a  power-law contribution.
The wavefunction renormalization factor of the field $i$ is
\begin{equation}
Z_i=1-\frac{(\gamma_i(\mu_0)-\tilde \gamma_i(\mu_0))}{2\pi}\ln\frac{\Lambda}{\mu_0}-
\frac{X_{\delta}}{\delta}\tilde \gamma_i(\mu_0)\left[\left(\frac{\Lambda}{\mu_0}\right)^{\delta}-1\right]
\label{anomdim}
\end{equation}
leading, for the Higgs fields to
\bea
\tilde\gamma_{H_u}&=&-\frac{3}{10}\alpha_1-\frac{3}{2}\alpha_2 \\
\tilde\gamma_{H_d}&=&-\frac{3}{10}\alpha_1-\frac{3}{2}\alpha_2 
\eea
with $\alpha_i\equiv g_i^2/4\pi$.
For what the Yukawa couplings are concerned, we get
\begin{eqnarray}
16\pi^2\frac{d}{dt}y_t&=&y_t\Bigg[ 3|y_t|^2+X_\delta\left(
\frac{\Lambda}{\mu_0}\right)^\delta\Big(3|y_t|^2+|y_b|^2-\frac{16}{3}g_3^2\nonumber\\
&&-3g_2^2-\frac{13}{15}g_1^2\Big)\Bigg]   \\
16\pi^2\frac{d}{dt}y_b&=&y_b\Bigg[ 3|y_b|^2+|y_\tau|^2+X_\delta\left(\frac{\Lambda}{\mu_0}\right)
^\delta\Big(|y_t|^2+3|y_b|^2 \nonumber\\
&&-\frac{16}{3}g_3^2-3g_2^2-\frac{7}{15}g_1^2\Big)\Bigg]  \\
16\pi^2\frac{d}{dt}y_\tau &=&y_\tau\Bigg[
|y_\tau|^2+3|y_b|^2\nonumber\\
&&+X_\delta\left(\frac{\Lambda}{\mu_0}\right)^\delta\Big(3|y_\tau|^2-3g_2^2-\frac{9}{5}g_1^2\Big)\Bigg]
\end{eqnarray}
Finally we analyze the (supersymmetric) Higgs mass 
\begin{eqnarray}
16\pi^2\frac{d}{dt}\mu &=&\mu\Bigg[\left( 3|y_t|^2+3|y_b|^2+|y_{\tau}|^2\right) \\
&&+X_\delta\left(\frac{\Lambda}{\mu_0}\right)^\delta\left(-3g_2^2-\frac{3}{5}g_1^2\right)\Bigg]
 \nonumber
\end{eqnarray}
In the Higgs $N=2$ case $\mu$ will not be renormalized as a consequence of a non-renormalization theorem.
\subsubsection{Soft terms in the minimal scenario}
\label{sec:SoftTerms}
While the effective potential method is powerful in determining the runnings for the parameters
in the SUSY Lagrangian, it 
is difficult to apply it to the computation of the soft terms in presence of the KK states. 
This is why we have decided to also carry on an analysis in terms of Feynman diagrams. 
In Appendix A we then collect the general form of the running of the couplings and masses appearing
in (\ref{scalarpot}).
In Appendix B we collect all the contributions to the renormalization of the couplings and masses
in (\ref{supersoft}). As a check, the interested reader can sum (\ref{b1})-(\ref{b4}) to recover
the result of (\ref{a5}) when specialized to the parameter $b$ in (\ref{supersoft}).
The computation of the effect of extra dimensions is now straightforward: each time that in a Feynman
diagram KK states are allowed to circulate in the loop (external states can only be zero modes), 
the contribution of the diagram must be added to the $\tilde b$ coefficient in (\ref{runnings}).
We finally remind the reader that in an orbifold compactification the KK momentum is conserved only in the 
bulk. The wavefunction in the compactified dimension must, in fact, be expanded in a basis which is 
invariant under the discrete group acting on the compactified dimension. 
Furthermore the orbifold fixed points break  translation invariance along the extra dimensions.

Following the previous prescription we obtain the $\beta$-functions for the trilinear couplings 
\begin{eqnarray}
16\pi^2\frac{d}{dt}a_t&=&9a_t |y_t|^2+X_{\delta}(\frac{\Lambda}{Q_0})^{\delta}\Bigg\lbrace a_t
\left[ 9|y_t|^2+|y_b|^2-\frac{16}{3}g_3^2-3g_2^2-\frac{13}{15}g_1^2\right]\label{triatsoft} \nonumber\\
&&+2a_by_b^*y_t+y_t\left[\frac{32}{3}g_3^2M_3+6g_2^2M_2+\frac{26}{15}g_1^2M_1\right]\Bigg\rbrace  \\ 
16\pi^2\frac{d}{dt}a_b&=& 9a_b |y_b|^2+a_b |y_{\tau}|^2+2a_{\tau}y_{\tau}^*y_b+X_{\delta}(\frac{\Lambda}{Q_0})^{\delta}\Bigg\lbrace a_b\Big[ 9|y_b|^2+|y_t|^2+ \nonumber\\
&&-\frac{16}{3}g_3^2-3g_2^2-\frac{7}{15}g_1^2\Big]+2a_ty_t^*y_b+ \nonumber\\
&&+y_b\left[\frac{32}{3}g_3^2M_3+6g_2^2M_2+\frac{14}{15}g_1^2M_1\right]\Bigg\rbrace  
\end{eqnarray}
\begin{eqnarray}
16\pi^2\frac{d}{dt}a_\tau &=&3a_{\tau}|y_b|^2+3a_{\tau}|y_{\tau}|^2+X_{\delta}(\frac{\Lambda}{Q_0})^{\delta}\Bigg\lbrace a_\tau\Big[ 9|y_\tau|^2+ \nonumber\\
&&-3g_2^2-\frac{9}{5}g_1^2\Big]+6a_by_b^*y_\tau+y_\tau\left[6g_2^2M_2+\frac{18}{5}g_1^2M_1\right]\Bigg\rbrace 
\label{tricousoft}
\end{eqnarray}
The $\beta$-function for the $b$ parameter (crucial for the electroweak symmetry breaking mechanism), is given by
\begin{eqnarray}
16\pi^2\frac{d}{dt}b&=& b\left( 3|y_t|^2+3|y_b|^2+|y_\tau|^2\right)+\mu\left[6a_ty_t^*+6a_by_b^*+2a_\tau y_\tau^* \right] \\
&&+X_{\delta}\left(\frac{\Lambda}{Q_0}\right)^{\delta}\Bigg\lbrace
b\left(-3g_2^2-\frac{3}{5}g_1^2\right)+\mu\left(+6g_2^2M_2+\frac{6}{5}g_1^2M_1\right)\Bigg\rbrace \nonumber
\end{eqnarray}
Now we can show the $\beta$-function for all the soft squared masses. For the two Higgs mass parameters we have
\begin{eqnarray}
16\pi^2\frac{d}{dt}m^2_{H_u}&=& 6|y_t|^2\left(m^2_{H_u}+m^2_t+m^2_{\bar t}\right)+6|a_t|^2+\frac{3}{5}g_1^2 tr'(Ym^2) \\
&&+X_{\delta}\left(\frac{\Lambda}{Q_0}\right)^{\delta}\left\lbrace -6g_2^2|M_2|^2-\frac{6}{5}g_1^2|M_1|^2+\frac{3}{5}g_1^2\left(m^2_{H_u}-m^2_{H_d}\right)\right\rbrace \nonumber\\
16\pi^2\frac{d}{dt}m^2_{H_d}&=& 6|y_b|^2\left(m^2_{H_d}+m^2_t+m^2_{\bar b}\right)+6|a_b|^2+2|a_{\tau}|^2 \\
&&+2|y_{\tau}|^2\left(m^2_{H_d}+m^2_L+m^2_{\bar\tau}\right)-\frac{3}{5}g_1^2 tr'(Ym^2) \nonumber\\
&&+X_{\delta}\left(\frac{\Lambda}{Q_0}\right)^{\delta}\left\lbrace
-6g_2^2|M_2|^2-\frac{6}{5}g_1^2|M_1|^2-\frac{3}{5}g_1^2\left(m^2_{H_u}-m^2_{H_d}\right)
\right\rbrace  \nonumber
\end{eqnarray}
while the equations for the squarks mass terms read
\begin{eqnarray}
16\pi^2\frac{d}{dt}m^2_{t}&=&2|y_t|^2m^2_{\bar t}+2|y_b|^2m^2_{\bar b}+\frac{1}{5}g_1^2 tr'(Ym^2)  \\
&&+X_{\delta}\left(\frac{\Lambda}{Q_0}\right)^{\delta}\Bigg\lbrace -\frac{32}{3}g_3^2|M_3|^2-6g_2^2|M_2|^2-\frac{2}{15}g_1^2|M_1|^2+2|a_t|^2+2|a_b|^2  \nonumber\\
&&+2|y_t|^2\left(m^2_{H_u}+m^2_t\right)+2|y_b|^2\left(m^2_{H_d}+m^2_t\right)+\frac{1}{5}g_1^2\left(m^2_{H_u}-m^2_{H_d}\right)\Bigg\rbrace \nonumber\\
16\pi^2\frac{d}{dt}m^2_{\bar t}&=&4|y_t|^2m^2_t-\frac{4}{5}g_1^2 tr'(Ym^2)  \\
&&+X_{\delta}\left(\frac{\Lambda}{Q_0}\right)^{\delta}\Bigg\lbrace -\frac{32}{3}g_3^2|M_3|^2-\frac{32}{15}g_1^2|M_1|^2+4|a_t|^2  \nonumber\\
&&+4|y_t|^2\left(m^2_{H_u}+m^2_{\bar t}\right)-\frac{4}{5}g_1^2\left(m^2_{H_u}-m^2_{H_d}\right)\Bigg\rbrace \nonumber
\end{eqnarray}
\begin{eqnarray}
16\pi^2\frac{d}{dt}m^2_{\bar b}&=&4|y_b|^2m^2_t+\frac{2}{5}g_1^2 tr'(Ym^2)  \\
&&+X_{\delta}\left(\frac{\Lambda}{Q_0}\right)^{\delta}\Bigg\lbrace -\frac{32}{3}g_3^2|M_3|^2-\frac{8}{15}g_1^2|M_1|^2+4|a_b|^2  \nonumber\\
&&+4|y_b|^2\left(m^2_{H_d}+m^2_{\bar b}\right)+\frac{2}{5}g_1^2\left(m^2_{H_u}-m^2_{H_d}\right)\Bigg\rbrace \nonumber
\end{eqnarray}
Finally, for the sleptons masses we have
\bea
16\pi^2\frac{d}{dt}m^2_{L}&=&2|y_{\tau}|^2m^2_{\bar\tau}-\frac{3}{5}g_1^2 tr'(Ym^2)  \\
&&+X_{\delta}\left(\frac{\Lambda}{Q_0}\right)^{\delta}\Bigg\lbrace -6g_2^2|M_2|^2-\frac{6}{5}g_1^2|M_1|^2+2|a_{\tau}|^2  \nonumber\\
&&+2|y_{\tau}|^2\left(m^2_{H_d}+m^2_L\right)-\frac{3}{5}g_1^2\left(m^2_{H_u}-m^2_{H_d}\right)\Bigg\rbrace \nonumber\\
16\pi^2\frac{d}{dt}m^2_{\bar \tau}&=&4|y_{\tau}|^2m^2_{L}+\frac{6}{5}g_1^2 tr'(Ym^2)  \label{slepmasssoft}\\
&&+X_{\delta}\left(\frac{\Lambda}{Q_0}\right)^{\delta}\Bigg\lbrace -\frac{24}{5}g_1^2|M_1|^2+4|a_{\tau}|^2  \nonumber\\
&&+4|y_{\tau}|^2\left(m^2_{H_d}+m^2_{\bar\tau}\right)+\frac{6}{5}g_1^2\left(m^2_{H_u}-m^2_{H_d}\right)\Bigg\rbrace  \nonumber
\eea
where, in order to avoid cumbersome expressions, we have introduced the quantity
\be\label{primedTrace}
\textit{tr}'(Ym^2)\equiv \textit{tr}(Ym^2)-m^2_{H_u}+m^2_{H_d} 
\ee
where the usual trace over the hypercharge, defined in the MSSM, is given by
\bea
\textit{tr}(Ym^2) & = & m^2_{H_u}-m^2_{H_d}+m^2_Q-2\left(m^2_{\bar u}+m^2_{\bar c}+m^2_{\bar t}\right)\\
& & +m^2_{\bar d}+m^2_{\bar s}+m^2_{\bar b}-m^2_L+m^2_{\bar e}+m^2_{\bar\mu}+m^2_{\bar\tau}\nonumber
\eea
In our computations we have assumed the first two families to be degenerate in mass, such that
\bea
\begin{aligned}
m^2_Q&= 2m^2_{Q_1}+m^2_{Q_3} ~ & \\
m^2_L&= 2m^2_{L_1}+m^2_{L_3} ~ & \\
m^2_{\bar e}&= m^2_{\bar\mu}~ & \\
m^2_{\bar u}&= m^2_{\bar c} ~ & 
\end{aligned}
\eea

\subsubsection{The non minimal scenario}
\label{subsec:nonminimalN=1}
In the non minimal scenario we introduce KK towers also for the matter fields.
We must distinguish among three different cases: $\eta=1$ in which only the third family have a KK tower, 
$\eta=2$ in which only the first two families have a KK tower and $\eta=3$ in which all the families have 
KK tower. The choice for the $\eta=2$ case is dictated by the usual third family approximation and by a 
constraint imposed by the ISASUGRA code concerning the mass degeneracy of the first two families.

One of the consequences of having KK towers for the matter fields is that they no longer live at the 
orbifold fixed points and they are allowed to be in the bulk. 
 
The anomalous dimensions of the Higgs fields are the same in the $\eta=0$ and $\eta=2$ case. 
For $\eta=1, 3$ we find
\begin{eqnarray}
\tilde\gamma_{H_u}&=&\gamma_{H_u} \\
\tilde\gamma_{H_d}&=&\gamma_{H_d} 
\end{eqnarray} 
For the chiral fields $t, b, \tau$ we have a similar 
situation and the coefficients of the $\beta$-function have only a power-law contribution. 

Let us now discuss the soft terms: for the trilinear couplings and the $b$ parameter, we must distinguish 
between the $\eta=2$ and $\eta=1,3$ cases. In the former case the running is the same of the  
minimal case due to the third family approximation. 
In the latter case, once again, the $\beta$-functions coefficients have only a power law behaviour. 

Finally the running of soft scalar masses in the non minimal cases 
is given by a  logarithmic  term multiplied by
\be
\textit{tr}'(Ym^2)
\ee
and a power law term given by
\be
\textit{tr}(Ym^2)-\textit{tr}'(Ym^2)
\ee
In the $\eta=1$ case the primed trace is defined as
\be
\textit{tr}'_{\eta=1}(Ym^2) = 2m^2_{Q_1}-2\left(m^2_{\bar u}+m^2_{\bar c}\right)+m^2_{\bar d}+m^2_{\bar s}-
2m^2_{L_1}+2m^2_{\bar e}
\label{traceeta1}\ee
while for $\eta=3$
\be
\textit{tr}'_{\eta=3}(Ym^2)=0
\label{traceeta3}\ee
because all the families have KK tower and there is no logarithmic contribution.
Finally for the case $\eta=2$ we have
\be
\textit{tr}'_{\eta=2}(Ym^2)  = m^2_{Q_3}-2\left(m^2_{\bar t}\right)+m^2_{\bar b}-m^2_{L_3}+m^2_{\bar \tau}
\label{traceeta2}\ee

\subsection{Higgs N=2}
With respect to the previous Section, in this case the changes are given by the presence of the term
(\ref{eq:n=2intterms}).
Following~\cite{Dienes:1998vg} the wave function renormalization of the Higgs fields are given by
\bea
\tilde\gamma_{H_u}&=&0  \\
\tilde\gamma_{H_d}&=&0 
\eea
The one-loop beta functions for the Yukawa couplings are then
\begin{eqnarray}
16\pi^2\frac{d}{dt}y_t&=&y_t\Bigg[ 3|y_t|^2-\frac{3}{10}g_1^2-\frac{3}{2}g_2^2+X_\delta\left(\frac{\Lambda}
{Q_0}\right)^\delta\Big(3|y_t|^2+|y_b|^2-\frac{16}{3}g_3^2\nonumber\\
&&-\frac{3}{2}g_2^2-\frac{17}{30}g_1^2\Big)\Bigg]   \\
16\pi^2\frac{d}{dt}y_b&=&y_b\Bigg[ 3|y_b|^2+|y_\tau|^2-\frac{3}{10}g_1^2-\frac{3}{2}g_2^2+X_\delta
\left(\frac{\Lambda}{Q_0}\right)^\delta\Big(|y_t|^2+3|y_b|^2 \nonumber\\
&&-\frac{16}{3}g_3^2-\frac{3}{2}g_2^2-\frac{1}{6}g_1^2\Big)\Bigg]  \\
16\pi^2\frac{d}{dt}y_\tau &=&y_\tau\Bigg[
|y_\tau|^2+3|y_b|^2-\frac{3}{10}g_1^2-\frac{3}{2}g_2^2+X_\delta\left(\frac{\Lambda}{Q_0}\right)^\delta
\Big(3|y_\tau|^2\nonumber\\
&&-\frac{3}{2}g_2^2-\frac{3}{2}g_1^2\Big)\Bigg]
\end{eqnarray}
For the (supersymmetric) Higgs mass, in virtue of the non-renormalization theorem, we have
\be
\frac{d}{dt}\mu=0 
\ee
This equation holds for any $\eta$. The $\beta$-functions for the $b$ parameter reads
\begin{eqnarray}
16\pi^2\frac{d}{dt}b&=& b\left( 3|y_t|^2+3|y_b|^2+|y_\tau|^2\right)+
\mu\left(6a_ty_t^*+6a_by_b^*+2a_\tau y_\tau^* \right) \nonumber\\
&&+
b\left(-3g_2^2-\frac{3}{5}g_1^2\right)+\mu\left(+6g_2^2M_2+\frac{6}{5}g_1^2M_1\right) 
\end{eqnarray}

\subsubsection{Soft terms in the minimal scenario}
We begin with the trilinear terms, which look like
\begin{eqnarray}
16\pi^2\frac{d}{dt}a_t&=&a_t\left( 9|y_t|^2-\frac{3}{10}g_1^2-\frac{3}{2}g_2^2\right)+
X_{\delta}(\frac{\Lambda}{Q_0})^{\delta}\Bigg\lbrace a_t\left[ -\frac{16}{3}g_3^2-\frac{3}{2}g_2^2-
\frac{17}{30}g_1^2\right] \nonumber\\
&&+a_t\left(9|y_t|^2+|y_b|^2\right)+2a_by_b^*y_t \nonumber\\
&&+y_t\left[\frac{32}{3}g_3^2M_3+6g_2^2M_2+\frac{26}{15}g_1^2M_1\right]\Bigg\rbrace\label{triatn=2}\\
16\pi^2\frac{d}{dt}a_b&=& a_b\left(9|y_b|^2+|y_{\tau}|^2-\frac{3}{10}g_1^2-\frac{3}{2}g_2^2\right)+
2a_{\tau}y_{\tau}^*y_b\nonumber\\
&&+X_{\delta}(\frac{\Lambda}{Q_0})^{\delta}\Bigg\lbrace a_b\Big[ 9|y_b|^2+|y_t|^2-\frac{16}{3}g_3^2-
\frac{3}{2}g_2^2-\frac{1}{6}g_1^2\Big]+2a_ty_t^*y_b \nonumber\\
&&+y_b\left[\frac{32}{3}g_3^2M_3+6g_2^2M_2+\frac{14}{15}g_1^2M_1\right]\Bigg\rbrace  \\
16\pi^2\frac{d}{dt}a_\tau &=&a_{\tau}\left(3|y_b|^2+3|y_{\tau}|^2-\frac{3}{10}g_1^2-\frac{3}{2}g_2^2
\right)+6a_{b}y_b^*y_{\tau}\nonumber\\
&&+X_{\delta}(\frac{\Lambda}{Q_0})^{\delta}\Bigg\lbrace a_\tau\Big[ 9|y_\tau|^2-\frac{3}{2}g_2^2-
\frac{3}{2}g_1^2\Big]\nonumber\\
&&+y_\tau\left[6g_2^2M_2+\frac{18}{5}g_1^2M_1\right]\Bigg\rbrace 
\end{eqnarray}

The $\beta$-functions for the soft squared masses of the Higgs fields are
\begin{eqnarray}
16\pi^2\frac{d}{dt}m^2_{H_u}&=& 6|y_t|^2\left(m^2_{H_u}+m^2_t+m^2_{\bar t}\right)+6|a_t|^2+\frac{3}{5}g_1^2 
\textit{tr}(Ym^2) \\
&& -6g_2^2|M_2|^2-\frac{6}{5}g_1^2|M_1|^2 \nonumber\\
16\pi^2\frac{d}{dt}m^2_{H_d}&=& 6|y_b|^2\left(m^2_{H_d}+m^2_t+m^2_{\bar b}\right)+6|a_b|^2+2|a_{\tau}|^2 \\
&&+2|y_{\tau}|^2\left(m^2_{H_d}+m^2_L+m^2_{\bar\tau}\right)-\frac{3}{5}g_1^2 \textit{tr}(Ym^2) \nonumber\\
&&-6g_2^2|M_2|^2-\frac{6}{5}g_1^2|M_1|^2  \nonumber
\end{eqnarray}
while for the squarks we have 
\begin{eqnarray}
16\pi^2\frac{d}{dt}m^2_{t}&=&2|y_t|^2m^2_{\bar t}+2|y_b|^2m^2_{\bar b}+\frac{1}{5}g_1^2 \textit{tr}'(Ym^2)  
\label{prima}\\
&&+X_{\delta}\left(\frac{\Lambda}{Q_0}\right)^{\delta}\Bigg\lbrace -\frac{32}{3}g_3^2|M_3|^2-6g_2^2|M_2|^2-\frac{2}{15}g_1^2|M_1|^2+2|a_t|^2+2|a_b|^2  \nonumber\\
&&+2|y_t|^2\left(m^2_{H_u}+m^2_t\right)+2|y_b|^2\left(m^2_{H_d}+m^2_t\right)+\frac{1}{5}g_1^2\left(m^2_{H_u}-m^2_{H_d}\right)\Bigg\rbrace \nonumber\\
16\pi^2\frac{d}{dt}m^2_{\bar t}&=&4|y_t|^2m^2_t-\frac{4}{5}g_1^2 \textit{tr}'(Ym^2)  \\
&&+X_{\delta}\left(\frac{\Lambda}{Q_0}\right)^{\delta}\Bigg\lbrace -\frac{32}{3}g_3^2|M_3|^2-\frac{32}{15}g_1^2|M_1|^2+4|a_t|^2  \nonumber\\
&&+4|y_t|^2\left(m^2_{H_u}+m^2_{\bar t}\right)-\frac{4}{5}g_1^2\left(m^2_{H_u}-m^2_{H_d}\right)\Bigg\rbrace \nonumber\\
16\pi^2\frac{d}{dt}m^2_{\bar b}&=&4|y_b|^2m^2_t+\frac{2}{5}g_1^2 \textit{tr}'(Ym^2)  \\
&&+X_{\delta}\left(\frac{\Lambda}{Q_0}\right)^{\delta}\Bigg\lbrace -\frac{32}{3}g_3^2|M_3|^2-\frac{8}{15}g_1^2|M_1|^2+4|a_b|^2  \nonumber\\
&&+4|y_b|^2\left(m^2_{H_d}+m^2_{\bar b}\right)+\frac{2}{5}g_1^2\left(m^2_{H_u}-m^2_{H_d}\right)\Bigg\rbrace \nonumber
\end{eqnarray}
Finally the equations for the leptons read
\bea
16\pi^2\frac{d}{dt}m^2_{L}&=&2|y_{\tau}|^2m^2_{\bar\tau}-\frac{3}{5}g_1^2 \textit{tr}'(Ym^2)  \\
&&+X_{\delta}\left(\frac{\Lambda}{Q_0}\right)^{\delta}\Bigg\lbrace -6g_2^2|M_2|^2-\frac{6}{5}g_1^2|M_1|^2+2|a_{\tau}|^2  \nonumber\\
&&+2|y_{\tau}|^2\left(m^2_{H_d}+m^2_L\right)-\frac{3}{5}g_1^2\left(m^2_{H_u}-m^2_{H_d}\right)\Bigg\rbrace \nonumber\\
16\pi^2\frac{d}{dt}m^2_{\bar \tau}&=&4|y_{\tau}|^2m^2_{L}+\frac{6}{5}g_1^2 \textit{tr}'(Ym^2)  \label{m2taun2}\\
&&+X_{\delta}\left(\frac{\Lambda}{Q_0}\right)^{\delta}\Bigg\lbrace -\frac{24}{5}g_1^2|M_1|^2+4|a_{\tau}|^2  \nonumber\\
&&+4|y_{\tau}|^2\left(m^2_{H_d}+m^2_{\bar\tau}\right)+\frac{6}{5}g_1^2\left(m^2_{H_u}-m^2_{H_d}\right)\Bigg\rbrace \nonumber
\eea
Our results (\ref{prima})-(\ref{m2taun2}) for the masses of the squarks and sleptons differ from those 
reported in \cite{Kobayashi:1998ye}
in two respects: on the one hand the terms proportional to $Y$ and $m^2_{H_u}-m^2_{H_d}$, due to the $U(1)$ 
gauge factor \footnote{For an explicit evaluation of this term, given by the diagram (\ref{diagzou2}),
see for example  (4.12) in \cite{Yamada:1994id}.},  are absent in \cite{Kobayashi:1998ye}. On the other hand 
in (\ref{prima})-(\ref{m2taun2}) there are terms coming from the diagrams (\ref{diagzou1}) which
do not get contributions from the KK states and therefore do not have a power-law type running. 

\subsubsection{Non minimal scenario}
The results we found in Section \ref{subsec:nonminimalN=1} for $N=1$ are still valid.
The only difference concerns the soft mass terms:
for the Higgs masses, $m^2_{H_u}, m^2_{H_d}$  no diagrams will contribute to the power law. 
The equations are then the same as in the MSSM. On the other hand the squark masses have only power 
law contributions.
For the other soft terms the equations are the same of the minimal scenario, regardless of the value 
of $\eta$.

\section{Phenomenology}
\setcounter{equation}{0}
\label{sec:pheno}
In this Section we present the results for the low energy phenomenology of our scenario.

To compute the weak scale parameters we used the publicly available codes ISASUGRA~\cite{Paige:2003mg} 
and DarkSUSY~\cite{Gondolo:2004sc}. We modified the ISAJET routines in order to take into account the 
power-law running of the gauge couplings, of the Yukawa couplings and of all the soft terms following 
the procedure outlined in the previous Sections. 
We solved the differential equations given by the running of the couplings and masses imposing  
boundary conditions as in the usual mSUGRA scenario~\cite{msugra}. In this way we have, at the new unification scale 
$M'_{GUT}$, a universal scalar mass $m_0$, a universal gaugino mass $m_{1/2}$ and a common trilinear 
coupling $A_0$. The 
only other parameters which need to be specified are $\tan\beta$ and $sign(\mu)$.

For a fixed choice of the extra-dimensional parameters $\mu_0$, $\delta$, $\eta$ we performed a detailed 
scan of the ($m_0$,$m_{1/2}$) parameter space keeping fixed $\tan\beta$, $A_0$ and $sign(\mu)$. 
We computed for every model the thermal relic density 
\be
\Omega_\chi h^2=\frac{m_\chi n_\chi}{\rho_c}
\ee
solving the Boltzmann equation for the neutralino number density $n_\chi(t)$
\be\label{Boltzeqcopy}
\frac{dn_\chi}{dt}+3Hn_\chi=-\langle\sigma_A v\rangle\left[(n)^2-(n_{eq})^2\right]
\ee
where $H=\dot{a}/a$ is the Hubble expansion parameter (with dot denoting time derivative), $a=a(t)$ is the 
scale factor of the Universe, $\langle\sigma_A v\rangle$ is the thermal averaged annihilation cross section 
and $n_{eq}$ is the equilibrium number density.
We considered only tree-level annihilation processes in the computation of $\langle\sigma_A v\rangle$. 
This implies that we do not have to worry about the presence of the KK states. In fact 
KK modes are too heavy to be produced in the intermediate tree-level annihilation 
channels. This is guaranteed by the fact that the neutralino mass (i.e. the mass of the lightest 
supersymmetric particle) is at most of order of $1$ TeV for every model we considered. 
For the relic density computation we have taken into account all the possible coannihilations with other 
sparticles. Because the mass of the first KK excited level is of order $\mu_0$ there is no possible 
coannihilation between the neutralino (a 0-mode particle) and the KK particles. In this way the standard 
treatment of the coannihilations is still valid.
The possible difference with the standard scenario may arise from the Hubble expansion rate $H$. In the 
standard framework, this term is responsible for the density $n_\chi\propto a^{-3}$, in absence of 
interactions.
Unfortunately in the literature there is not a general expression of the Hubble expansion rate $H$ for an 
arbitrary number of extra dimensions 
$\delta$.\footnote{See \cite{Maartens:2003tw,Kiritsis:2003mc,Quevedo:2002xw} for general reviews on 
the subject.}
There are some interesting issues concerning the $D=6$ 
dimensional case in~\cite{Kofinas:2004ae, Kofinas:2005py} and also 
in~\cite{Multamaki:2003zc, Cuadros-Melgar:2005ex}. We considered explicitly only the case $D=5$~\cite{Apostolopoulos:2004ic} and we 
argued that the conclusion still holds in the higher dimensional case \cite{Kiritsis1}.
Let us consider the case $\delta=1$. Following~\cite{Langlois:2002bb} and~\cite{Langlois:2003yx} we have
\be
H^2=\frac{8\pi G}{3}\rho\left(1+\frac{\rho}{\rho_0}\right)+\frac{{\cal C}}{a^4}
\ee
where 
\be
\rho_0=96\pi G M_5^6
\ee
with $M_5$ being the five dimensional Planck mass. The third term, where ${\cal C}$ is an 
integration constant, is referred to as the dark radiation term. The second term proportional to $\rho^2$ 
and the dark radiation term are the new ingredients in the brane world 
cosmology and they lead to a non standard expansion law. 
The second term is the dominant one when $\rho\gg\rho_0$. In order to satisfy this condition we must have $M_5\lesssim 10^4$ TeV.

The $D=5$ Planck mass is related to the $D=4$ Planck mass through the relation 
\be
(M_5)^3R=M^2_P
\ee 
where $R\equiv 1/\mu_0$ is the compactification scale. 
In our model we obtain the values collected in Table~\ref{mu0M5rho5}. This result holds in general in model with flat extra space-time dimensions. Typical values of the halo density $\rho$ are of order of $1\: {\rm GeV}/{\rm cm}^3$ so the $\rho^2$ term is always subdominant. Hence the usual Friedmann equation holds.
\begin{table}[ht]
\begin{center}
\begin{tabular}{|c|c|c|}
\hline
$\mu_0\equiv 1/R$ & $M_5=(\mu_0 M_P^2)^{1/3}$ & $\rho_0=96\pi G M_5^6$ \\
(Gev) & (GeV) & (${\rm GeV/cm^3}$)\\
\hline
\hline
$10^5$ & $2.15\cdot 10^{14}$ & $ 10^{8}$ \\
$10^6$ & $4.64\cdot 10^{14}$ & $ 10^{10}$ \\
$10^8$ & $2.15\cdot 10^{15}$ & $ 10^{14}$ \\
$10^{10}$ & $1\cdot 10^{16}$ & $ 10^{18}$ \\
$10^{12}$ & $4.64\cdot 10^{16}$ & $10^{22}$ \\
\hline
\end{tabular}
\caption{Values of the $D=5$ Planck mass $M_5$ and of $\rho_0$ for different values of the compactification scale $\mu_0$.}
\label{mu0M5rho5}
\end{center}
\end{table}

Let us fix the notation defining the lightest neutralino as the linear combination
\be
\tilde{\chi}_1^0=N_{11}\tilde{B}+N_{12}\tilde{W}+N_{13}\tilde{H}_u+N_{14}\tilde{H}_d
\ee 
where $\tilde{B}$ and $\tilde{W}$ are the bino and wino fields while $\tilde{H}_u$ and $\tilde{H}_d$ are the two higgsinos. We also define the gaugino fraction as
\be
Z_g=\left|N_{11}\right|^2+\left|N_{12}\right|^2
\ee
We say that a neutralino is gaugino-like (in particular in our case bino-like) if $Z_g>0.9$ while is higgsino-like when $Z_g<0.1$. In all the intermediate cases we denote the neutralino as mixed-like.


As explained in Sec.\ref{sec:running} we considered two different scenarios: one in which the two zero-mode Higgs fields are chiral $N=1$ superfields as in the MSSM and the other in which they form an $N=2$ matter hypermultiplet.
 
\begin{figure}[ht]
\begin{center}
\includegraphics[scale=0.7]{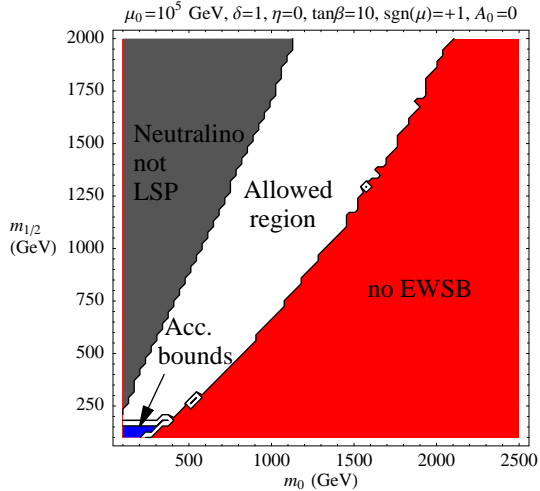}
\caption{Excluded regions in the plane ($m_0$, $m_{1/2}$)}
\label{exclmu0105}
\end{center}
\end{figure}

\begin{figure}[ht]
\begin{center}
\includegraphics[scale=0.7]{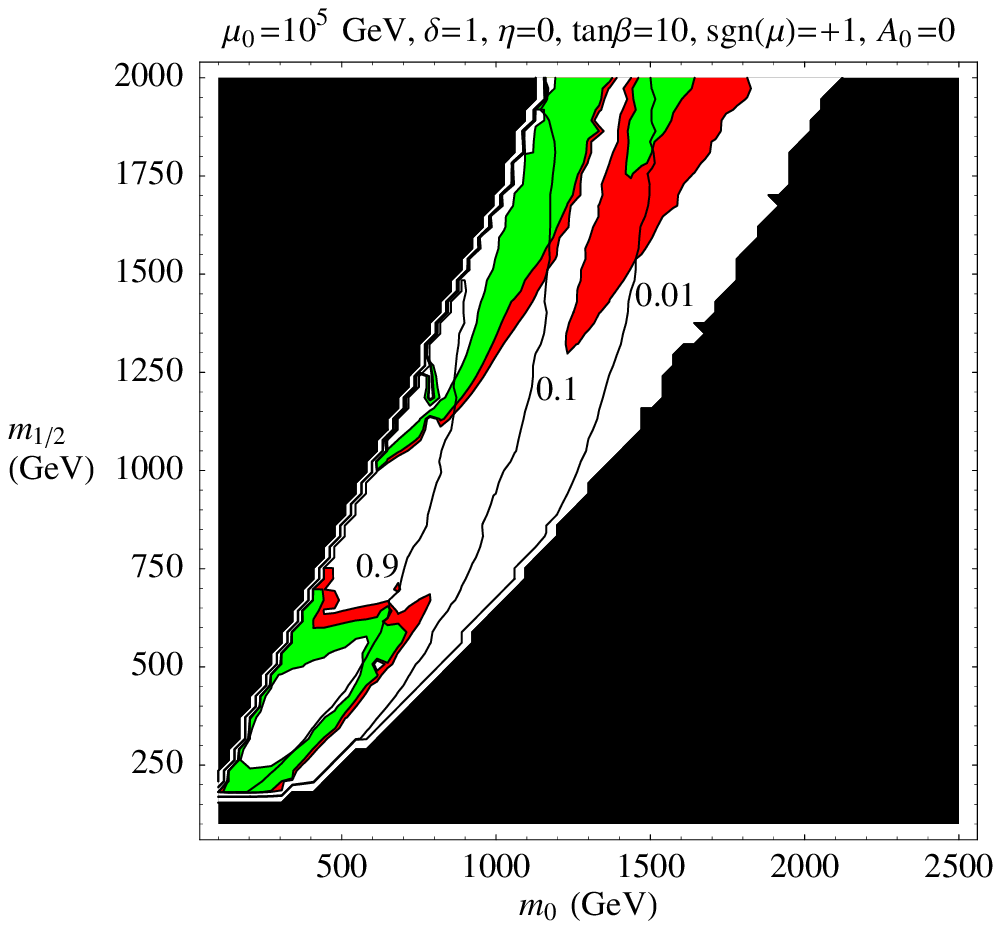}
\includegraphics[scale=0.7]{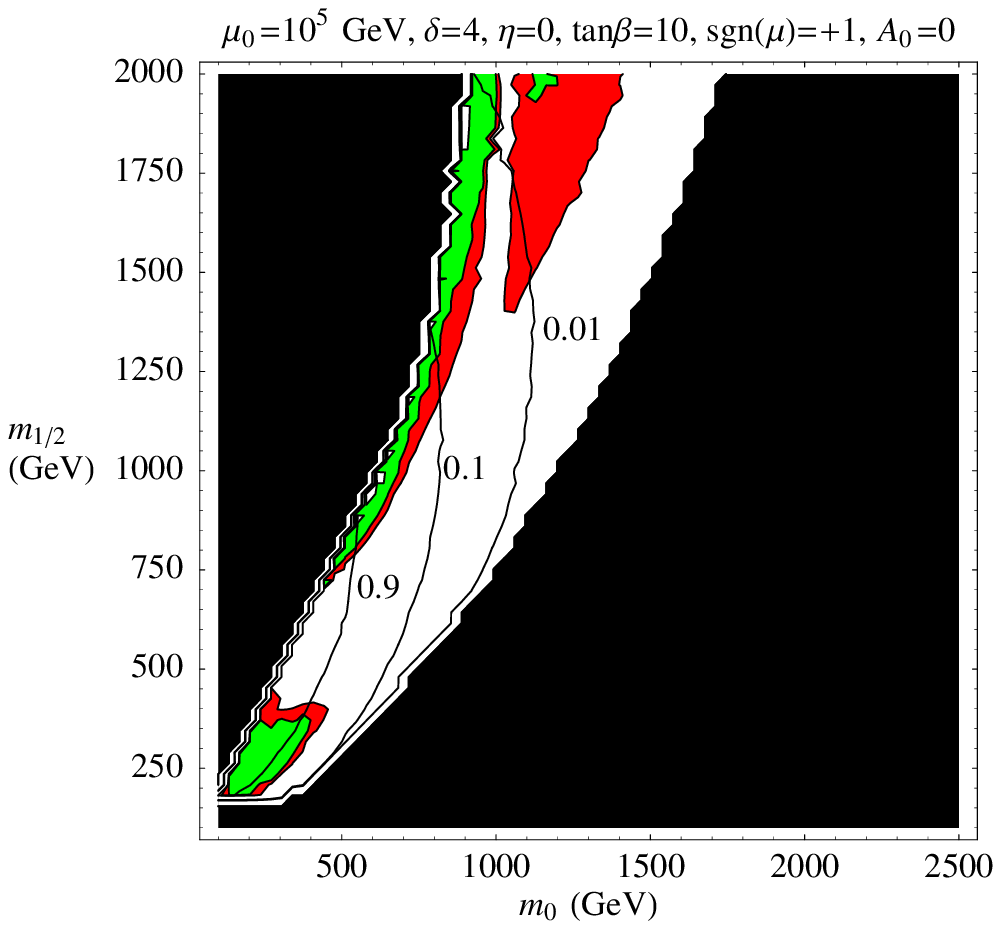}
\caption{Contour plots of the gaugino fraction in the plane ($m_0$, $m_{1/2}$) for different values of $\delta$.}
\label{zgmu0105}
\end{center}
\end{figure}

We first analyze the $N=1$ Higgs case in the minimal scenario $\eta=0$ in which the matter fermions do not have KK towers. 
We performed a detailed scan in the parameter space ($m_0$, $m_{1/2}$) fixing all the other parameters. In Fig.~\ref{exclmu0105} we show the regions already excluded due to either theoretical or experimental reasons for a given model ($\mu_0=10^5$ GeV, $\delta=1$ and $\eta=0$). The dark gray region is excluded because the $\tilde{\tau}$ is the LSP rather than the neutralino, the red region is excluded because the models do not achieve electroweak symmetry breaking (EWSB) while the blue region is excluded because the models do not satisfy the current accelerator bounds (limits on the chargino masses, $b\to s\gamma$, etc.).
In almost all the parameter space the neutralino is still the LSP. One of the main result is that, unlike the standard mSUGRA case, the neutralino is no longer bino-like but it tends to be a very pure higgsino 
(see Fig.~\ref{zgmu0105}).
This conclusion strongly depends from the value of the compactification scale $\mu_0$. In fact the regions in which the neutralino is higgsino-like get smaller for higher values of $\mu_0$ (see the left panel of Fig.~\ref{zgmu0105} for the case $\mu_0=10^5$ GeV and Fig.~\ref{mu0105vs108} for the two cases $\mu_0=10^8$ GeV and $\mu_0=10^{10}$ GeV). For higher values of $\mu_0$ the higgsino-like region approaches the region excluded due to an incorrect EWSB.
The previous result is only slightly dependent from the number of extra-dimensions $\delta$ as can be 
seen by 
comparing the two panels of Fig.~\ref{zgmu0105}. The only difference is a change in the shape of the 
excluded regions.
In other words the crucial property of the $\beta$-functions is the power-law behaviour rather than the effective 
power-law index, i.e. $\delta$.

In every contour plot for the gaugino fraction we have shown the cosmologically allowed regions. The 
red regions are those for which the relic density $\Omega h^2$ satisfies the WMAP 
constraints \cite{Spergel:2003cb,Tegmark:2003ud}
\begin{equation}
0.09\le\Omega h^2\le0.13
\end{equation}
while the green regions denote the pre-WMAP constraints
\begin{equation}
0.13<\Omega h^2\le 0.30
\end{equation}
The cosmologically allowed regions have a huge overlap with the pure higgsino-like region, especially for low values of $\mu_0$, i.e. $\mu_0\lesssim 10^8$ GeV. For higher values of $\mu_0$ the cosmologically allowed regions tend in general to overlap with a mixed-like neutralino region.

\begin{figure}[ht]
\begin{center}
\includegraphics[scale=0.7]{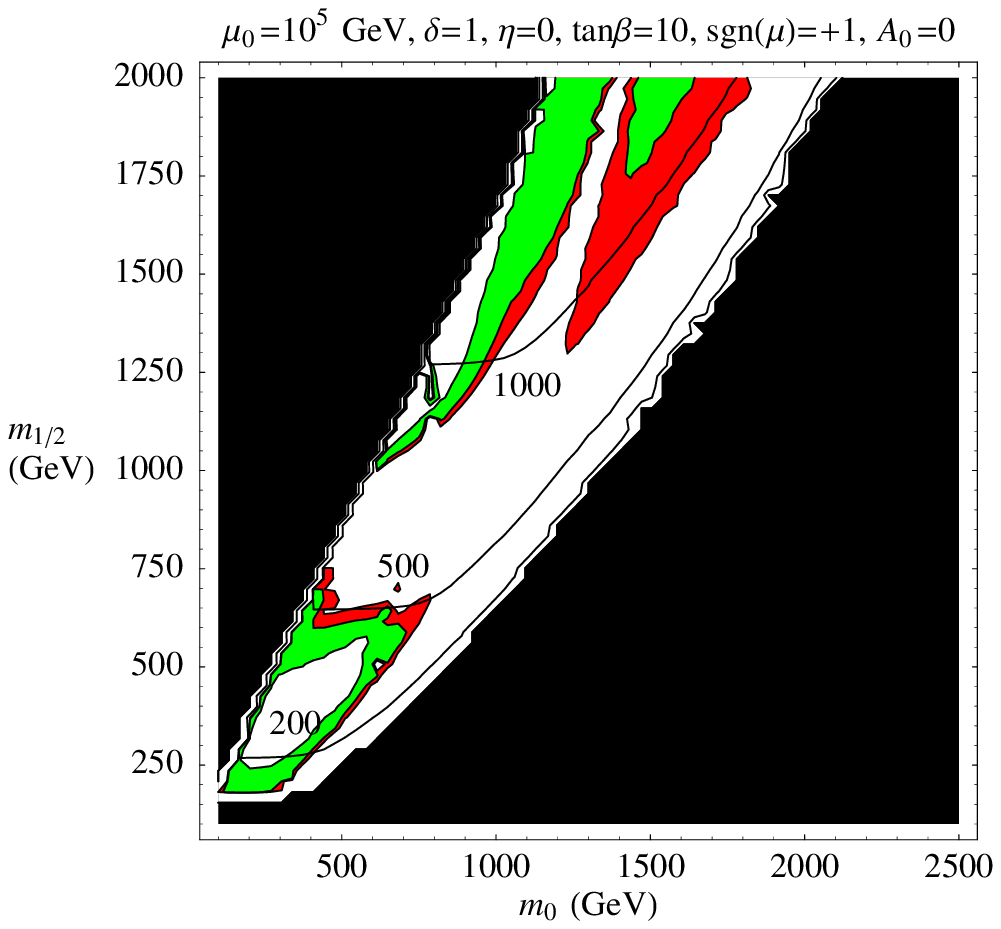}
\includegraphics[scale=0.7]{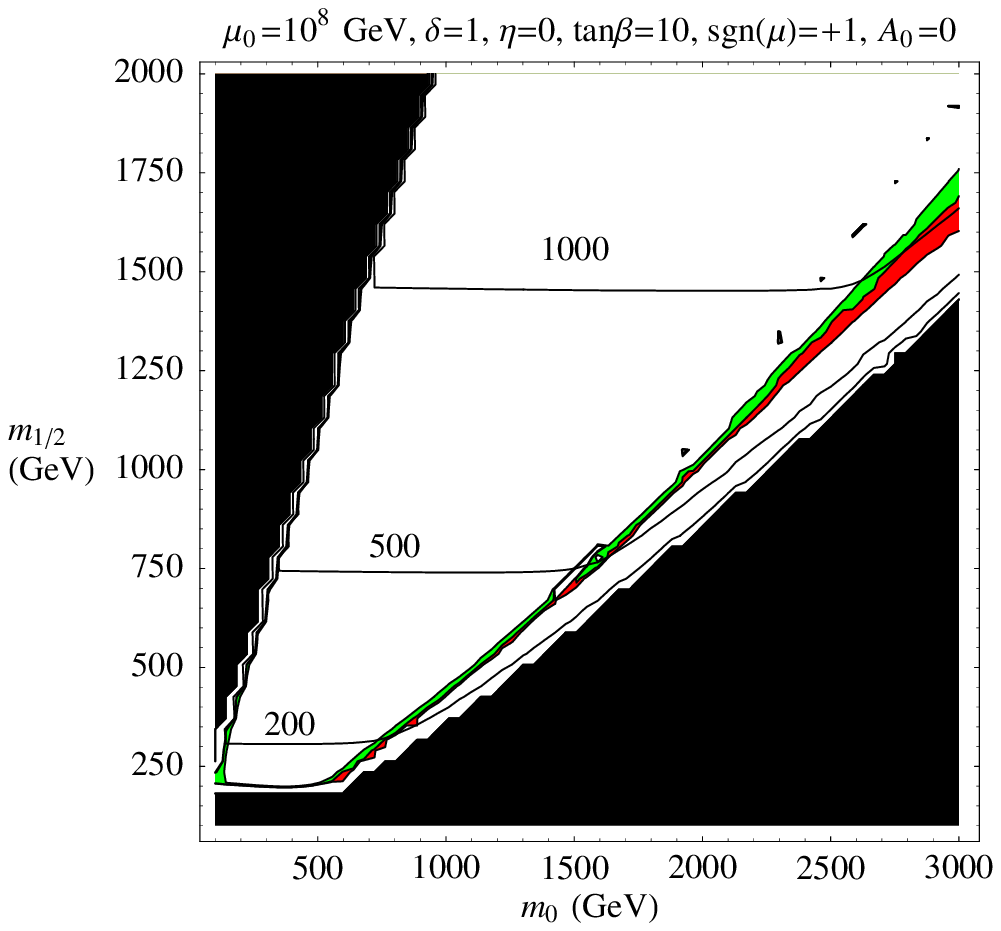}
\includegraphics[scale=0.7]{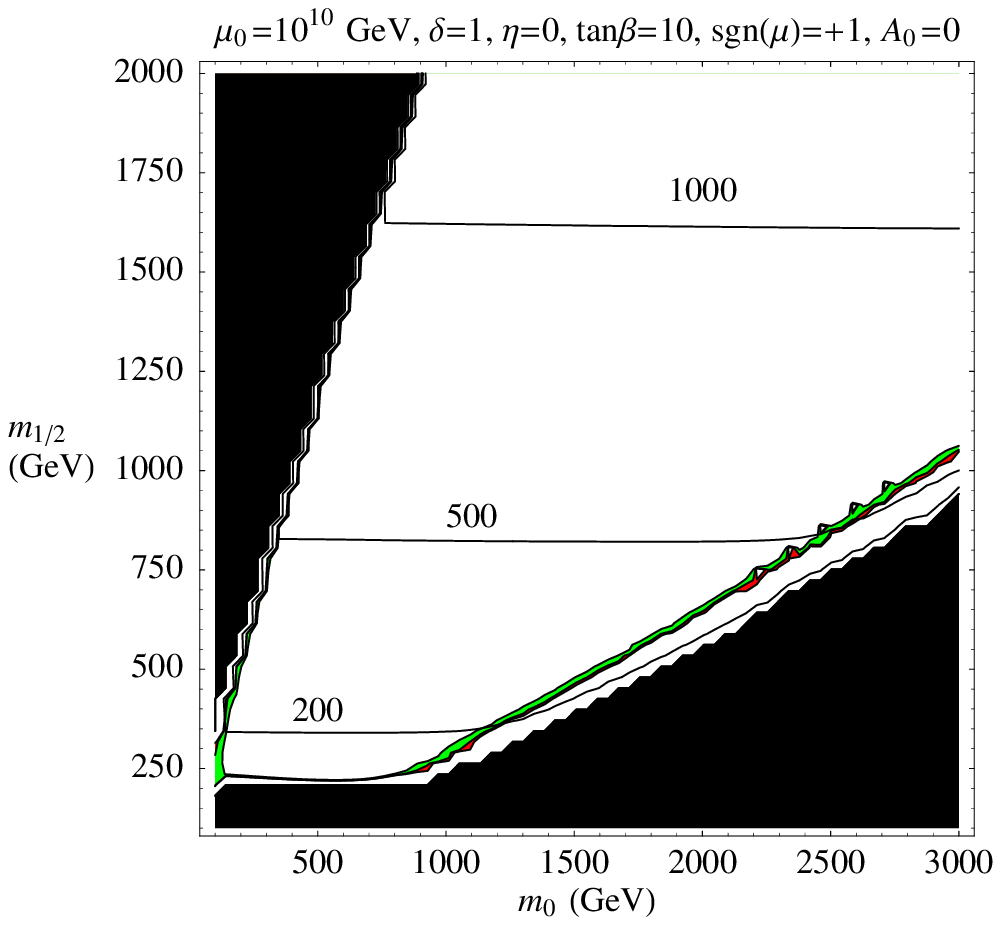}
\includegraphics[scale=0.7]{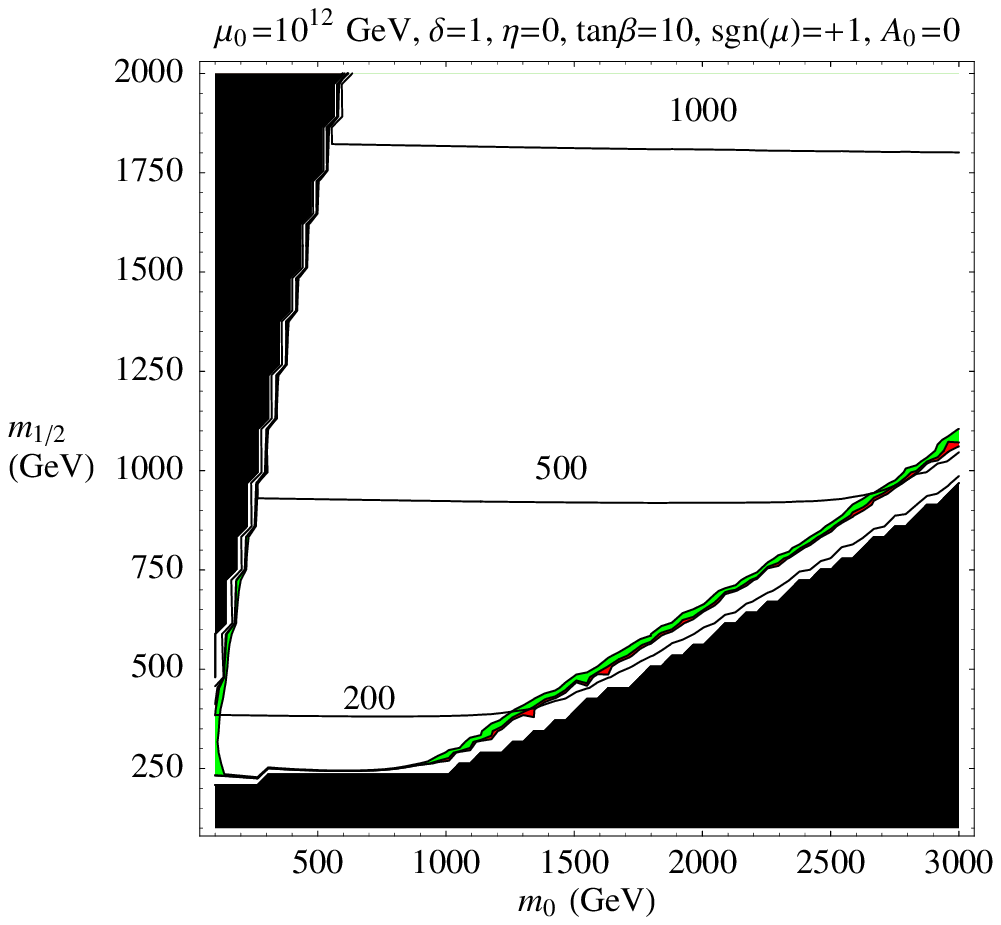}
\caption{Contour plots of the neutralino mass (in GeV) in the plane ($m_0$, $m_{1/2}$) for increasing values of $\mu_0$.}
\label{mchi_increase_mu0}
\end{center}
\end{figure}

\begin{figure}[ht]
\begin{center}
\includegraphics[scale=0.7]{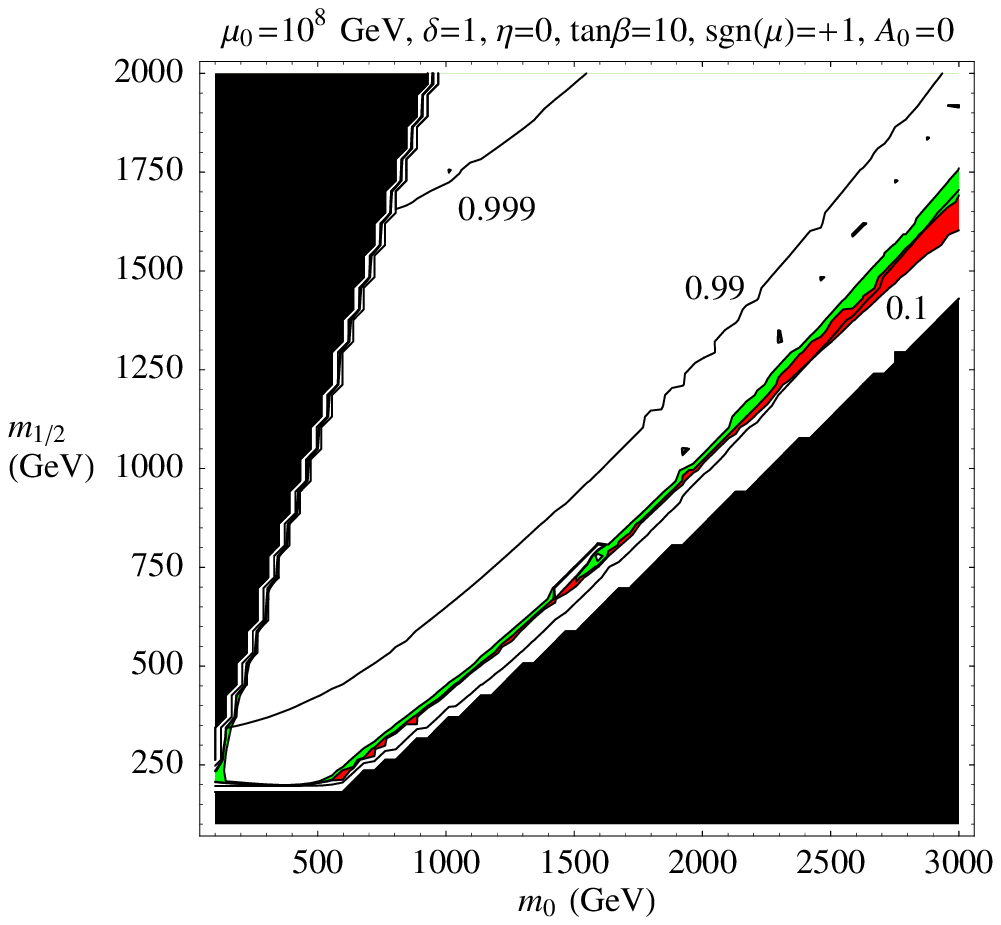}
\includegraphics[scale=0.7]{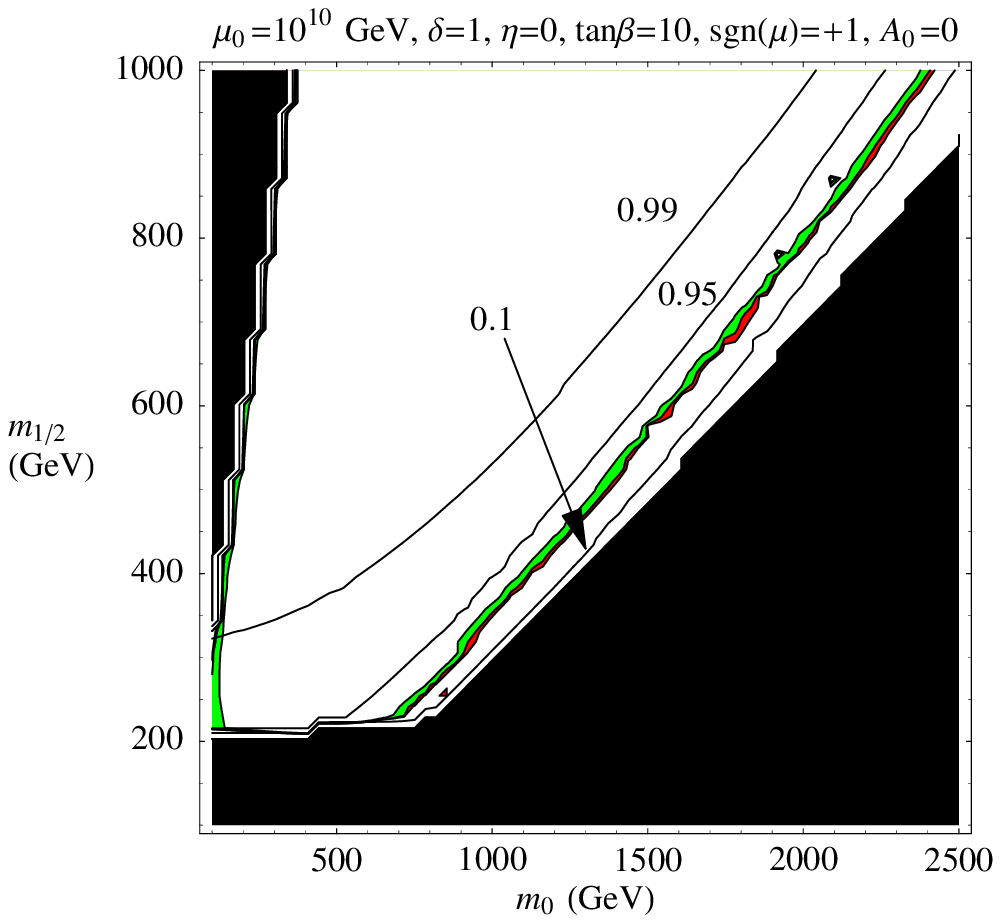}
\caption{Contour plots of the gaugino fraction in the plane ($m_0$, $m_{1/2}$) for increasing values of $\mu_0$.}
\label{mu0105vs108}
\end{center}
\end{figure}

We also present in Fig.~\ref{mchi_increase_mu0} the isomass contour plots for the neutralino for increasing $\mu_0$ and for fixed $\delta$. As in the case of the gaugino fraction we show the cosmologically allowed regions. For higher values of $\mu_0$ high neutralino mass contours (around 1 TeV) are shifted upwards so that models that possess the right relic density have $m_\chi\lesssim 500$ GeV.
Increasing the value of $A_0$, for example $A_0= 2500$, implies a shift of the cosmologically allowed regions towards higher values of the neutralino mass of about $m_\chi\gtrsim 800$ GeV and a wider region excluded by the accelerator bounds. Moreover, in this case, the neutralino is a very pure higghsino $Z_g<0.1$ in all the allowed parameter space.

For increasing values of $\tan\beta$ we obtain the same shift of the cosmologically allowed regions as in the case of high $A_0$. Moreover the excluded region in which the neutralino is not the LSP grows towards higher values of $m_0$.

Let us now analyze the non minimal models with $\eta>0$.
As explained in subsection~\ref{subsec:nonminimalN=1} we considered three different cases: $\eta=1$ in which only the third family have a KK tower, $\eta=2$ in which only the first two families have a KK tower and $\eta=3$ in which all the families have KK tower.

For $\eta=2$ we must have $\mu_0\ge 10^8$ GeV in order to avoid non perturbative gauge couplings at the unification scale (see for example Fig.~4 of \cite{Dienes:1998vg}). In this case regions with $m_0\gtrsim 3000$ GeV are excluded due to an incorrect EWSB while for $\tan\beta\lesssim 30$ there is no region in which the neutralino is not the LSP. For higher $\tan\beta$ the parameter space develop a region, for small $m_0$, excluded by EWSB constraints and by the stau being the LSP. In almost all the allowed region the neutralino turns out to be a very pure bino (as in mSUGRA) though there is still a significant overlap between the cosmologically favoured region and the region in which the neutralino is higgsino or mixed-like. 
The possible values of the neutralino mass are lower than in the minimal scenario, with typically $m_\chi\lesssim 500$ GeV.

In the case $\eta=1$ in which only the third family has KK tower almost all the parameter space is allowed for low $\tan\beta$. There are only two small regions excluded by the accelerator bounds and by the EWSB constraints. The neutralino turns out to be always a bino. The cosmologically allowed region is very small and in correspondence with a low neutralino mass region, i.e. $m_\chi\lesssim 50$ GeV. For higher $\tan\beta$ we have the same behaviour except for the presence of the two excluded regions (EWSB constraints and the stau being the LSP) for small $m_0$.

\begin{figure}[ht]
\begin{center}
\includegraphics[scale=0.7]{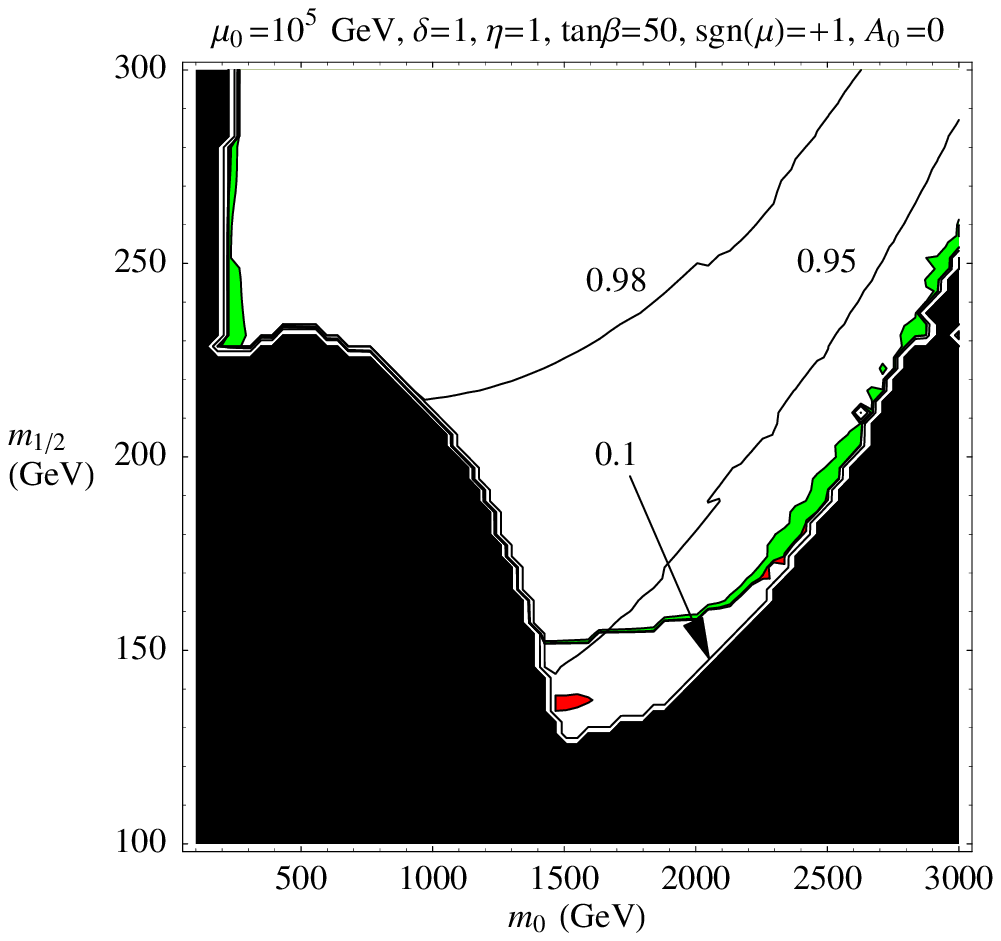}
\includegraphics[scale=0.7]{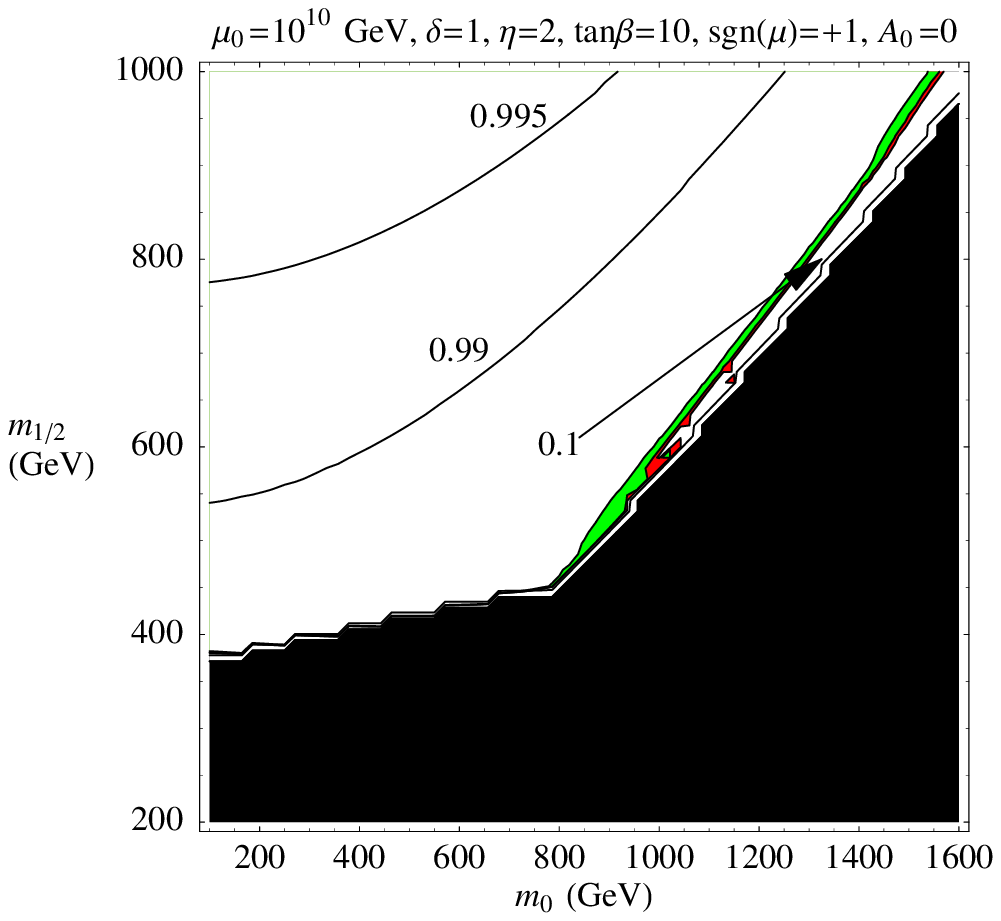}
\caption{Gaugino fraction in the plane ($m_0$, $m_{1/2}$). Left panel: non minimal scenario $\eta=1$. Right panel: non minimal scenario $\eta=2$.}
\label{zg_eta1_eta2}
\end{center}
\end{figure}

\begin{figure}[ht]
\begin{center}
\includegraphics[scale=0.7]{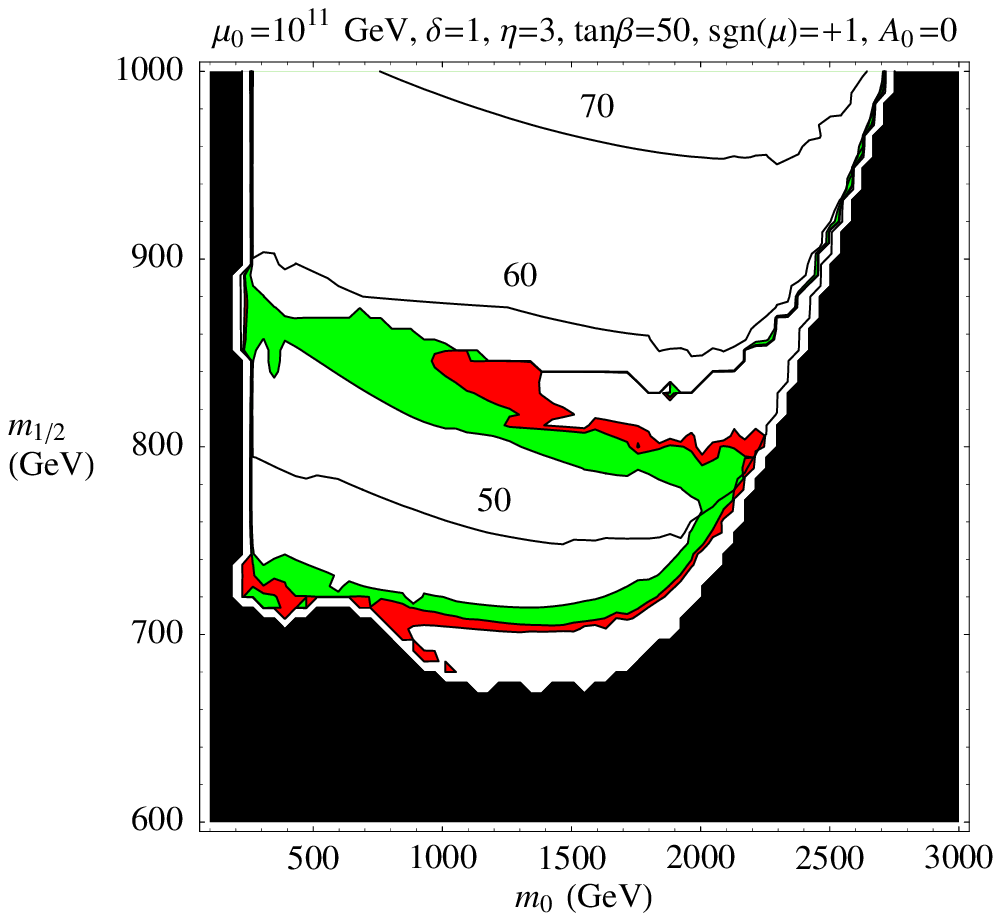}
\includegraphics[scale=0.7]{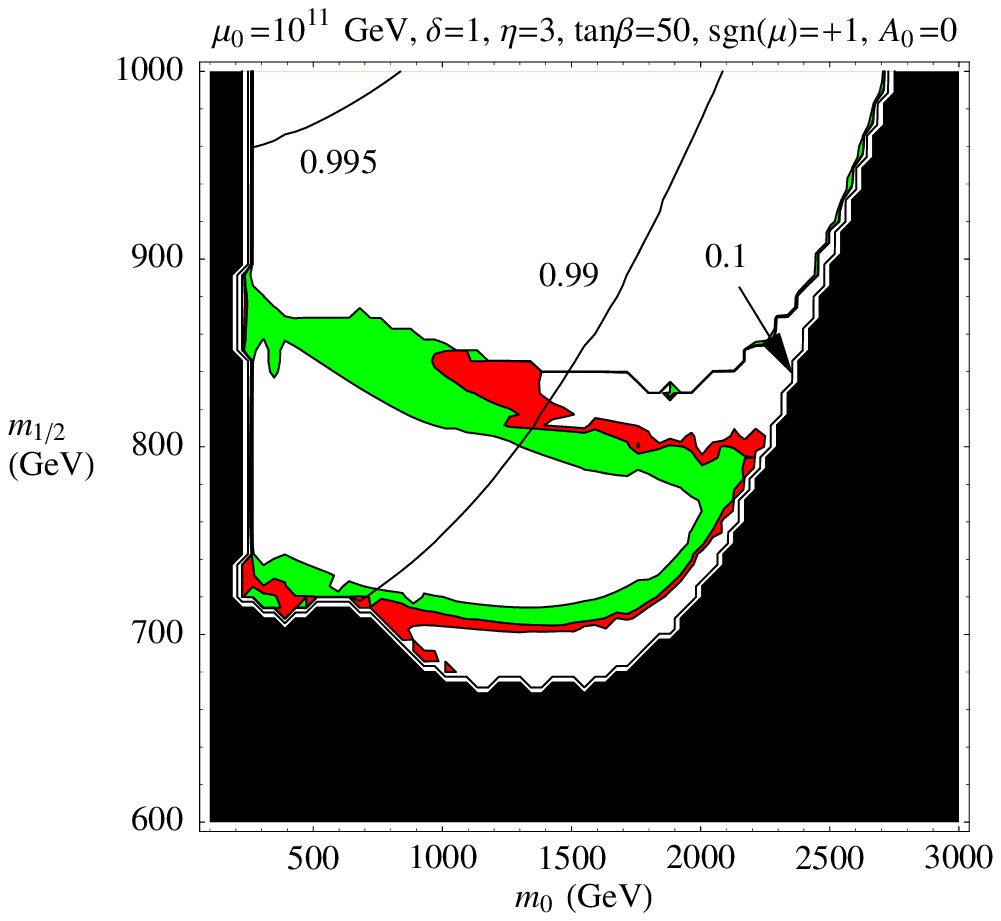}
\caption{Left panel: neutralino isomass contours ($m_0$, $m_{1/2}$) for the non minimal scenario $\eta=3$. Right panel: gaugino fraction in the plane ($m_0$, $m_{1/2}$) for the non minimal scenario $\eta=3$.}
\label{mchi_zg_eta3}
\end{center}
\end{figure}

Finally we consider the $\eta=3$ case in which all the families possess KK tower. In these models the region for which $m_{1/2}\lesssim 750$ GeV is always excluded for any value of $\tan\beta$. 
For $\tan\beta\lesssim 30$ there is only a very small region in which the neutralino is not the LSP, namely for $m_0\simeq 0$ GeV and $m_{1/2}\simeq 0$ GeV. For higher values of $\tan\beta$ there is another region in which the neutralino is not the LSP together with a region excluded by the EWSB on the left of the parameter space.
The neutralino mass for this kind of models is lower, $m_\chi\lesssim 200$ GeV , with respect to the minimal case (left panel of Fig.~\ref{mchi_zg_eta3}). The neutralino composition is essentially that of a pure bino except for a region close to the excluded regions in which $Z_g\le 0.1$ (right panel of Fig.~\ref{mchi_zg_eta3}). Once again the relic density is in the right range in the regions where the neutralino is higgsino or mixed-like.

In general non minimal models are disfavoured from the point of view of the cosmological relic density because the allowed regions are very small.

\begin{figure}
\centering
\subfigure[Excluded regions in the plane 
($m_0$, $m_{1/2}$) for the minimal scenario $\eta=0$.]{
\includegraphics[scale=0.7]{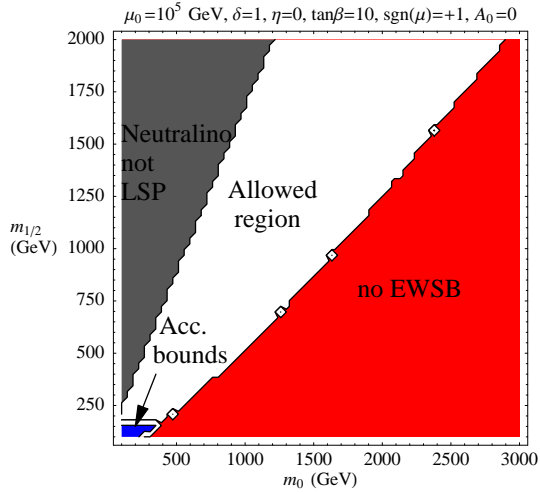}}
\subfigure[ Gaugino fraction in the plane 
($m_0$, $m_{1/2}$) for the minimal scenario $\eta=0$.]{
\includegraphics[scale=0.7]{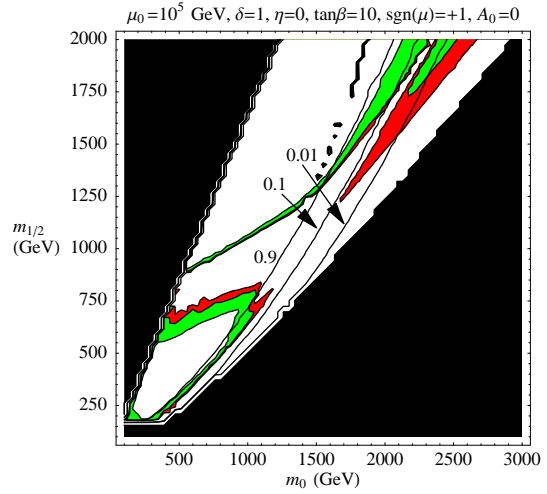}}
\subfigure[Neutralino isomass contour 
in the plane ($m_0$, $m_{1/2}$) for the non minimal scenario $\eta=1$.]{
\includegraphics[scale=0.7]{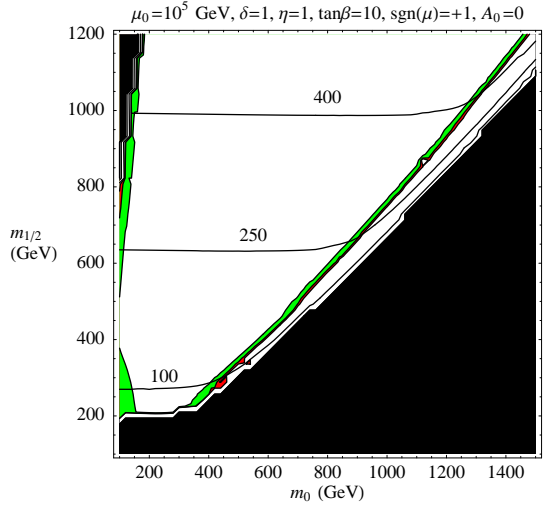}}
\subfigure[ Neutralino 
isomass contour in the plane ($m_0$, $m_{1/2}$) for the non minimal scenario $\eta=2$.]{
\includegraphics[scale=0.7]{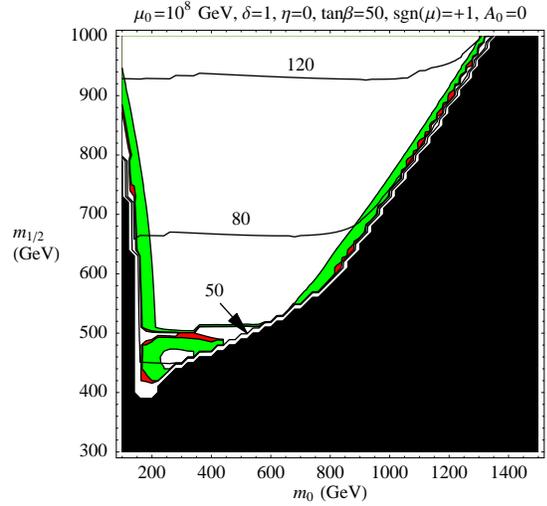}}
\caption{Contour plots for $N=2$ Higgs models.}
\label{excl_zg_n=2}
\end{figure}

In the scenario in which the Higgs form an $N=2$ matter hypermultiplet the results about the neutralino composition and the corresponding relic density remain essentially unchanged. 
However there are some differences in the shape of the allowed and excluded regions in the parameter space.
 For example in the minimal scenario $\eta=0$ (see Fig.~\ref{excl_zg_n=2}(a)) the regions 
excluded due to EWSB and to the accelerator bounds are smaller than in $N=1$ case. 
This holds for all $\tan\beta$ 
and $A_0$. 
Keeping fixed all the other parameters there are in general allowed regions in the right part of the 
parameter space with $m_0>2500$ GeV.
The cosmologically allowed regions get shifted in order to ``follow'' the regions in which $Z_g\lesssim 0.1$.
In the non minimal cases $\eta>0$ the behaviour is the opposite and the parameter space is in general more constrained, although there are cases in which the region excluded by the neutralino not being the 
LSP is absent or at least very small (see Fig.~\ref{excl_zg_n=2}(c),(d)).

\section{Conclusions}
In this paper we presented a general scenario that involves the presence of large extra space-time dimensions. We computed all the threshold corrections (due to KK modes) to the running of the couplings of the MSSM, including all the soft terms. Then we performed a detailed phenomenological analysis of the model.
We found that the neutralino is still the LSP with a relic density compatible with the most recent WMAP data. The main difference between our model and the well studied mSUGRA scenario is that the neutralino is higgsino like in most of the parameter space. Moreover the mass spectrum of the theory does not contain tachyons. 
These results remain essentially the same in the two cases we considered, $N=1$ and $N=2$ Higgs.

\section*{Acknowledgements}
The authors would like to thank M. Bianchi and E. Trevigne for useful discussions. 
This work was supported in part by the EC contracts MRTN-CT-2004-512194 and MRTN-CT-2004-503369, the MIUR-COFIN contract 2003-023852, the NATO contract PST.CLG.978785, the INTAS contracts
03-51-6346.

\newpage
\appendix
\setcounter{equation}{0}
\section{$\beta$-functions from the method of the effective potential}
In this appendix we collect the full set of renormalization group equations following the notation of 
\cite{derendinger}
\begin{eqnarray}
Q\frac{d}{dQ}f_{abc}&=&-\frac{1}{32\pi^2}\Big[4g^2[C(A)+C(B)+C(C)]f_{abc}\nonumber\\
&&-(X^{a'}_af_{a'bc}+X^{b'}_bf_{ab'c}+X^{c'}_cf_{abc'})\Big] \label{a1}\\
Q\frac{d}{dQ}\mu_{ab}&=&-\frac{1}{32\pi^2}\Big[4g^2[C(A)+C(B)]\mu_{ab}-(X^{a'}_a\mu_{a'b}+X^{b'}_b\mu_{ab'})\Big]  \label{a2}\\
Q\frac{d}{dQ}(m^2)^a_b&=&-\frac{1}{32\pi^2}\Big[-X^a_c(m^2)^c_b-X^c_b(m^2)^a_c-4 f^{ade}f_{bce}(m^2)^c_d+ \nonumber\\
&&-8\xi_{(3)acd}\xi^{bcd}_{(3)}+8g^2[C(A)\Delta_A^2+C(B)\Delta_B^2]\delta^a_b \nonumber\\
&&-4g^2tr(T^Am^2)(T^A)^a_b\Big] \label{a3}\\
Q\frac{d}{dQ}\xi_{(3)abc}&=&-\frac{1}{32\pi^2}\Big[+4g^2[C(A)+C(B)+C(C)]\xi_{(3)abc} \nonumber\\
&&-2\left(f_{bcd}f^{b'c'd}\xi_{(3)ab'c'}+f_{acd}f^{a'c'd}\xi_{(3)a'bc'}+f_{abd}f^{a'b'd}\xi_{(3)a'b'c}\right) \nonumber\\
&&- X^{a'}_a\xi_{(3)a'bc}-X^{b'}_b\xi_{(3)ab'c}-X^{c'}_c\xi_{(3)abc'} \nonumber\\
&&+4g^2[C(A)\Delta_A+C(B)\Delta_B+C(C)\Delta_C]f_{abc}\Big] \label{a4}\\
Q\frac{d}{dQ}\xi_{(2)ab}&=&-\frac{1}{32\pi^2}\Big[+4g^2[C(A)+C(B)]\xi_{(2)ab}-[X^{a'}_a\xi_{(2)a'b}+X^{b'}_b\xi_{(2)ab'}] \nonumber\\
&&-2f_{abc}f^{cde}\xi_{(2)de}-4f^{cde}[\xi_{(3)acd}\mu_{eb}+\xi_{(3)acd}\mu_{eb}]   \nonumber\\
&&+8g^2[C(A)\Delta_A+C(B)\Delta_B]\mu_{ab}\Big]  \label{a5}\\
Q\frac{d}{dQ}\xi_{(1)a}&=&-\frac{1}{32\pi^2}\Big[+4g^2C(A)\xi_{(1)a}-X^{b}_a\xi_{(1)b}-2f^{bcd}\mu_{ab}\xi_{(2)cd} \nonumber\\
&&-4\xi_{(3)abc}\xi_{(2)}^{bc}-4f_{abc}\mu^{cd}(m^2)^b_d\Big] \label{a6}
\end{eqnarray}

The generalization of these equations to non simple gauge groups
is very simple. Terms like $g^2[C(A)+C(B)]$ are to be
replaced by a sum over the gauge groups (all simple
components and $U(1)$ factors). There is a gauge coupling 
constant and a group invariant $C(R)$ for
each group factor.
\section{Feynman diagrams analysis of the soft terms}
\setcounter{equation}{0}
In this appendix we collect all the Feynman diagrams which renormalize 
the soft terms in (\ref{supersoft}).
Next to each diagram we give its contribution to the coefficient of the respective
$\beta$-function. As one can easily check, in the absence of KK states, every diagram is 
logarithmically divergent in four dimensions. 
In presence of extra dimensions, those diagrams in which KK states can circulate contribute
to the power law piece in (\ref{runnings}).
This is the recipe we have adopted to obtain (\ref{triatsoft})-(\ref{slepmasssoft})
and (\ref{triatn=2})-(\ref{m2taun2})
\begin{itemize}
\item Renormalization of the coefficient $b$ in (\ref{supersoft})
\end{itemize}
\begin{eqnarray}
\parbox{20mm}{
\begin{fmfgraph*}(100,80)
 \fmfpen{thin}
 \fmfleft{i}
 \fmfv{label=$H$,label.angle=60}{i}
 \fmfright{o}
 \fmfv{label=$H$,label.angle=120}{o}
 \fmf{dashes}{i,v4}
 \fmf{dashes}{v4,v1}
 \fmf{dashes}{v2,v3}
 \fmfv{decor.shape=cross,decoration.size=50}{v3}
 \fmf{dashes}{v3,o}
 \fmf{plain,left,tension=0.3,label=$t,,b,,\tau$}{v1,v2}
 \fmf{plain,right,tension=0.3}{v1,v2}
 \end{fmfgraph*}
}
&&\hspace{2cm} b(3 y_t^2+3y^2_b+y^2_\tau) \label{b1}\\
\parbox{20mm}{
\begin{fmfgraph*}(100,80)
 \fmfleft{i}
 \fmfv{label=$H$,label.angle=60}{i}
 \fmfright{o}
 \fmfv{label=$H$,label.angle=120}{o}
 \fmf{dashes}{i,v1}
 \fmf{dashes}{v2,o}
 \fmf{dashes,left,tension=0.5,label=$\tilde t,,\tilde b,,\tilde\tau $}{v1,v2}
 \fmf{dashes,right,tension=0.5}{v1,v2}
 \end{fmfgraph*}}
&&\hspace{2cm} 2\mu(3a_t y_t+3 a_b y_b+a_\tau y_\tau) \label{b2}\\
\parbox{20mm}{
\begin{fmfgraph*}(100,80)
 \fmfpen{thin}
 \fmfleft{i}
 \fmfv{label=$H$,label.angle=60}{i}
 \fmfright{o}
 \fmfv{label=$H$,label.angle=120}{o}
 \fmf{dashes}{i,v1}
 \fmf{dashes}{v1,v2}
 \fmfv{decor.shape=cross}{v2}
 \fmfv{label=$H$,label.angle=-90}{v2}
 \fmf{dashes}{v2,v3}
 \fmf{dashes}{v3,o}
 \fmf{photon,left,tension=0}{v1,v3}
\end{fmfgraph*}
}
&&\hspace{2cm} b(-3g^2_2-\frac{3}{5}g^2_1)\label{b3}\\
\parbox{20mm}{
\begin{fmfgraph*}(100,80)
 \fmfpen{thin}
 \fmfleft{i}
 \fmfv{label=$H$,label.angle=60}{i}
 \fmfright{o}
 \fmfv{label=$H$,label.angle=120}{o}
 \fmf{dashes}{i,v1}
 \fmf{dashes}{v3,o}
 \fmf{cross_gauge,left,tension=0.2}{v1,v3}
 \fmf{plain,left,tension=0.2,label=$\lambda$}{v1,v3}
 \fmf{plain,right,tension=0.6,label=$\tilde H$}{v1,v3}
 \end{fmfgraph*}
}
&& \hspace{2cm} -\mu(6g^2_2 \Delta^{22}+\frac{6}{5}g_1^2 \Delta^{11})\label{b4}
\end{eqnarray}
\newpage
\begin{itemize}
\item Renormalization of the coefficient $m^2_{H_u}, m^2_{H_d}$ in (\ref{supersoft})
\end{itemize}
\begin{eqnarray}
\parbox{30mm}{
\begin{fmfgraph*}(100,80)
 \fmfpen{thin}
 \fmfleft{i}
 \fmfv{label=$H$,label.angle=60}{i}
 \fmfright{o}
 \fmfv{label=$H$,label.angle=120}{o}
 \fmf{dashes}{i,v4}
 \fmf{dashes}{v4,v1}
 \fmf{dashes}{v2,v3}
 \fmfv{decor.shape=cross,decoration.size=50}{v3}
 \fmf{dashes}{v3,o}
 \fmf{plain,left,tension=0.3,label=$t,,b$}{v1,v2}
 \fmf{plain,right,tension=0.3}{v1,v2}
 \end{fmfgraph*}
}
&&\hspace{2cm} 6y_t^2 m^2_{H_u} \\
\parbox{30mm}{
\begin{fmfgraph*}(100,80)
 \fmfleft{i}
 \fmfv{label=$H$,label.angle=60}{i}
 \fmfright{o}
 \fmfv{label=$H$,label.angle=120}{o}
 \fmf{dashes}{i,v1}
 \fmf{dashes}{v2,o}
 \fmf{dashes,left,tension=0.5,label=$\tilde t$}{v1,v2}
 \fmf{dashes,right,tension=0.5}{v1,v2}
 \end{fmfgraph*}}
&&\hspace{2cm} 6 a_t^2 \\
\parbox{30mm}{
\begin{fmfgraph*}(100,80)
 \fmfpen{thin}
 \fmfleft{i}
 \fmfv{label=$H$,label.angle=60}{i}
 \fmfright{o}
 \fmfv{label=$H$,label.angle=120}{o}
 \fmf{dashes}{i,v1}
 \fmf{dashes}{v3,o}
 \fmf{doublecross_gaugethird,left,tension=0.2}{v1,v3}
 \fmf{doublecross_gaugetwothird,left,tension=0.2}{v1,v3}
 \fmf{wiggly,left,tension=0.2,label=$\lambda$}{v1,v3}
 \fmf{plain,right,tension=0.6,label=$\tilde H$}{v1,v3}
 \end{fmfgraph*}
}
&&\hspace{2cm} -\frac{6}{5}g_1^2(\Delta^{11})^2-6g_2^2(\Delta^{22})^2\\
\parbox{30mm}{
\begin{fmfgraph*}(100,80)
 \fmfpen{thin}
 \fmfleft{i}
 \fmfv{label=$H$,label.angle=60}{i}
 \fmfright{o}
 \fmfv{label=$H$,label.angle=120}{o}
 \fmf{dashes}{i,v}
 \fmf{dashes}{v,o}
 \fmf{cross_scalar,label=$\tilde t$}{v,v}
 \end{fmfgraph*}
}
&& \hspace{2cm} 6y_t^2(m^2_t+m^2_{\bar t}) \\
\parbox{30mm}{
\begin{fmfgraph*}(100,80)
 \fmfpen{thin}
 \fmfleft{i}
 \fmfv{label=$H$,label.angle=60}{i}
 \fmfright{o}
 \fmfv{label=$H$,label.angle=120}{o}
 \fmf{dashes}{i,v}
 \fmf{dashes}{v,o}
 \fmf{cross_scalar,label=$\tilde t,, H$}{v,v}
 \end{fmfgraph*}
} && \hspace{2cm} \frac{3}{5}g_1^2 \textit{tr}(Ym^2)
\end{eqnarray} 
The last diagram is present only for the $U(1)$ gauge group.
\newpage
\begin{itemize}
\item Renormalization of the coefficient $m^2_{\tilde t}$ in (\ref{supersoft})
\end{itemize}
\begin{eqnarray}
\parbox{20mm}{
\begin{fmfgraph*}(100,80)
 \fmfpen{thin}
 \fmfleft{i}
 \fmfv{label=$\tilde t$,label.angle=60}{i}
 \fmfright{o}
 \fmfv{label=$\tilde t$,label.angle=120}{o}
 \fmf{dashes}{i,v4}
 \fmf{dashes}{v4,v1}
 \fmf{dashes}{v2,v3}
 \fmfv{decor.shape=cross,decoration.size=50}{v3}
 \fmf{dashes}{v3,o}
 \fmf{plain,left,tension=0.3,label=$\tilde H$}{v1,v2}
 \fmf{plain,right,tension=0.3,label=$t$}{v1,v2}
 \end{fmfgraph*}
}
&&\hspace{2cm}  (2y_t^2+2y_b^2)m^2_{t}\\
\parbox{20mm}{
\begin{fmfgraph*}(100,80)
 \fmfleft{i}
 \fmfv{label=$\tilde t$,label.angle=60}{i}
 \fmfright{o}
 \fmfv{label=$\tilde t$,label.angle=120}{o}
 \fmf{dashes}{i,v1}
 \fmf{dashes}{v2,o}
 \fmf{dashes,left,tension=0.5,label=$H$}{v1,v2}
 \fmf{dashes,right,tension=0.5,label=$\tilde t$}{v1,v2}
 \end{fmfgraph*}}
&&\hspace{2cm} 2 (a_t^2+a_b^2) \\
\parbox{20mm}{
\begin{fmfgraph*}(100,80)
 \fmfpen{thin}
 \fmfleft{i}
 \fmfv{label=$\tilde t$,label.angle=60}{i}
 \fmfright{o}
 \fmfv{label=$\tilde t$,label.angle=120}{o}
 \fmf{dashes}{i,v1}
 \fmf{dashes}{v3,o}
 \fmf{doublecross_gaugethird,left,tension=0.2}{v1,v3}
 \fmf{doublecross_gaugetwothird,left,tension=0.2}{v1,v3}
 \fmf{wiggly,left,tension=0.2,label=$\lambda$}{v1,v3}
 \fmf{plain,right,tension=0.6,label=$t$}{v1,v3}
 \end{fmfgraph*}
}
&&\hspace{2cm} -8g^2\Delta^2_A C(t)\\
\parbox{20mm}{
\begin{fmfgraph*}(100,80)
 \fmfpen{thin}
 \fmfleft{i}
 \fmfv{label=$\tilde t$,label.angle=60}{i}
 \fmfright{o}
 \fmfv{label=$\tilde t$,label.angle=120}{o}
 \fmf{dashes}{i,v}
 \fmf{dashes}{v,o}
 \fmf{cross_scalar,label=$\tilde t,,H$}{v,v}
 \end{fmfgraph*}
}
&& \hspace{2cm} 2y_t^2(m^2_{H_u}+m^2_{\bar t})+ 2y_b^2(m^2_{H_d}+m^2_{\bar b})\label{diagzou1}\\
\parbox{20mm}{
\begin{fmfgraph*}(100,80)
 \fmfpen{thin}
 \fmfleft{i}
 \fmfv{label=$\tilde t$,label.angle=60}{i}
 \fmfright{o}
 \fmfv{label=$\tilde t$,label.angle=120}{o}
 \fmf{dashes}{i,v}
 \fmf{dashes}{v,o}
 \fmf{cross_scalar,label=$\tilde t,, H$}{v,v}
 \end{fmfgraph*}
} && \hspace{2cm} \frac{1}{5}g_1^2 \textit{tr}(Ym^2)\label{diagzou2}
\end{eqnarray}
Similar diagrams hold for the $\tilde b$ squark and the
$\tilde\tau$ slepton.
\newpage
\begin{itemize}
\item Renormalization of the coefficient $a_t, a_b, a_\tau$ in (\ref{supersoft})
\end{itemize}
\begin{eqnarray}
\parbox{20mm}{
\begin{fmfgraph*}(90,50)
 \fmfpen{thin}
 \fmfleft{i}
 \fmfv{label=$H$,label.angle=120}{i}
 \fmfright{o1,o2}
\fmffreeze
 \fmfv{label=$\tilde t$,label.angle=50}{o1}
 \fmfv{label=$\tilde t$,label.angle=-50}{o2}
 \fmf{dashes}{i,v1}
 \fmf{dashes}{v1,v2}
 \fmf{dashes}{v2,v3}
 \fmf{dashes}{v3,v4}
\fmf{dashes,tension=.4}{o1,v4}
\fmf{dashes,tension=.4}{o2,v4}
 \fmf{photon,left,tension=0}{v1,v3}
 \end{fmfgraph*}
}
&& \hspace{2cm} a_t(-\frac{3}{10}g_1^2-\frac{3}{2}g_2^2) \\ && \nonumber\\ && \nonumber \\
\parbox{20mm}{
\begin{fmfgraph*}(90,50)
 \fmfpen{thin}
 \fmfleft{i}
 \fmfv{label=$H$,label.angle=120}{i}
 \fmfright{o1,o2}
\fmffreeze
 \fmfv{label=$\tilde t$,label.angle=50}{o1}
 \fmfv{label=$\tilde t$,label.angle=-50}{o2}
 \fmf{dashes}{i,v1}
 \fmf{dashes}{v1,v2}
 \fmf{dashes}{v2,v3}
 \fmf{dashes}{v3,v4}
\fmf{dashes,tension=.4}{o1,v4}
\fmf{dashes,tension=.4}{o2,v4}
 \fmf{photon,left,tension=0}{v1,v3}
 \end{fmfgraph*}
}
&&\hspace{2cm} a_t(-\frac{17}{30}g_1^2-\frac{3}{2}g_2^2-\frac{16}{3}g_3^2) \\&& \nonumber\\ && \nonumber \\
\parbox{20mm}{
\begin{fmfgraph*}(90,50)
  \fmfpen{thin}
 \fmfleft{i}
 \fmfv{label=$\tilde t$,label.angle=60}{i}
 \fmfright{o1,o2}
 \fmfv{label=$\tilde t$,label.angle=50}{o1}
 \fmfv{label=$H$,label.angle=-50}{o2}
 \fmf{dashes}{i,v1}
 \fmf{dashes}{v2,v3}
 \fmf{dashes,tension=.4}{v3,o1}
 \fmf{dashes,tension=.4}{v3,o2}
 \fmf{plain,left,tension=0.3,label=$\tilde H$}{v1,v2}
 \fmf{plain,right,tension=0.3,label=$t$}{v1,v2}
 \end{fmfgraph*}
}
&&\hspace{2cm} a_t(3y_t^2+y_b^2)\\&& \nonumber\\ && \nonumber \\
\parbox{20mm}{
\begin{fmfgraph*}(90,50)
   \fmfpen{thin}
 \fmfleft{i}
 \fmfv{label=$H$,label.angle=60}{i}
 \fmfright{o1,o2}
 \fmfv{label=$H$,label.angle=50}{o1}
 \fmfv{label=$\tilde t$,label.angle=-50}{o2}
 \fmf{dashes}{i,v1}
 \fmf{dashes,label=$H$}{v2,v3}
 \fmf{dashes,tension=.4}{v3,o1}
 \fmf{dashes,tension=.4}{v3,o2}
 \fmf{plain,left,tension=0.3,label=$t,,b,,\tau$}{v1,v2}
 \fmf{plain,right,tension=0.3,label=$t,,b,,\tau$}{v1,v2}
 \end{fmfgraph*}
}
&& \hspace{2cm} 3a_ty_t^2\\&& \nonumber\\ && \nonumber \\
\parbox{20mm}{
\begin{fmfgraph*}(90,50)
 \fmfpen{thin}
 \fmfleft{i}
 \fmfv{label=$H$,label.angle=120}{i}
 \fmfright{o1,o2}
 \fmfv{label=$\tilde t$,label.angle=50}{o1}
 \fmfv{label=$\tilde t$,label.angle=-50}{o2}
 \fmf{dashes}{i,v1}
 \fmf{dashes,tension=.4}{v2,o1}
 \fmf{dashes,tension=.4}{v2,o2}
 \fmf{dashes,left,tension=0.4,label=$\tilde t,,\tilde b,,\tilde\tau$}{v1,v2}
 \fmf{dashes,right,tension=0.4}{v1,v2}
 \end{fmfgraph*}
} && \hspace{2cm}  6a_ty_t^2 \\&& \nonumber\\ && \nonumber \\
\parbox{20mm}{
\begin{fmfgraph*}(90,50)
 \fmfpen{thin}
 \fmfleft{i}
 \fmfv{label=$\tilde t$,label.angle=120}{i}
 \fmfright{o1,o2}
 \fmfv{label=$H$,label.angle=50}{o1}
  \fmfv{label=$\tilde t$,label.angle=-50}{o2}
 \fmf{dashes}{i,v1}
 \fmf{dashes,tension=.4}{v2,o1}
 \fmf{dashes,tension=.4}{v2,o2}
 \fmf{dashes,left,tension=0.4,label=$H$}{v1,v2}
 \fmf{dashes,right,tension=0.4,label=$\tilde t$}{v1,v2}
 \end{fmfgraph*}
} && \hspace{2cm}  6a_ty_t^2+2a_by_ty_b\\&& \nonumber\\ && \nonumber \\
\parbox{20mm}{
\begin{fmfgraph*}(90,50)
 \fmfpen{thin}
 \fmfleft{i}
 \fmfv{label=$H$,label.angle=120}{i}
 \fmfright{o1,o2}
 \fmfv{label=$\tilde t$,label.angle=50}{o1}
  \fmfv{label=$\tilde t$,label.angle=-50}{o2}
 \fmf{dashes}{i,v1}
 \fmf{plain,tension=.4}{v1,v2}
 \fmfv{label=$t$,label.angle=-160,l.d=.15w}{v2}
 \fmf{plain,tension=.4,label=$t$}{v1,v3}
 \fmf{cross_gauge,tension=.05}{v2,v3}
 \fmf{plain,tension=.05}{v2,v3}
 \fmf{dashes,tension=0.4}{v2,o1}
 \fmf{dashes,tension=0.4}{v3,o2}
 \end{fmfgraph*}
} && \hspace{2cm} -y_t\left(\frac{16}{3}g_3^2\Delta^{33}+3g_2^2\Delta^{22}+\frac{13}{15}g_1^2\Delta^{11}
\right)\\&& \nonumber\\ && \nonumber \\
\parbox{20mm}{
\begin{fmfgraph*}(90,50)
 \fmfpen{thin}
 \fmfleft{i}
 \fmfv{label=$\tilde t$,label.angle=120}{i}
 \fmfright{o1,o2}
 \fmfv{label=$H$,label.angle=50}{o1}
  \fmfv{label=$\tilde t$,label.angle=-50}{o2}
 \fmf{dashes}{i,v1}
 \fmf{plain,tension=.4}{v1,v2}
 \fmfv{label=$\tilde H$,label.angle=-160,l.d=.15w}{v2}
 \fmf{plain,tension=.4,label=$t$}{v1,v3}
 \fmf{cross_gauge,tension=.05}{v2,v3}
 \fmf{plain,tension=.05}{v2,v3}
 \fmf{dashes,tension=0.4}{v2,o1}
 \fmf{dashes,tension=0.4}{v3,o2}
 \end{fmfgraph*}
} && \hspace{2cm} -y_t\left(\frac{16}{3}g_3^2\Delta^{33}+3g_2^2\Delta^{22}+
\frac{13}{15}g_1^2\Delta^{11}\right)
\eea
\end{fmffile}
Similar diagrams hold for the other two
trilinear couplings, $a_b, a_\tau$,  the one for the squark $\tilde b$ and the one
for the slepton $\tilde\tau$.


\begin{thebibliography}{99}\frenchspacing
\bibitem{Antoniadis:1988jn}
  I.~Antoniadis, C.~Bachas, D.~C.~Lewellen and T.~N.~Tomaras,
  Phys.\ Lett.\ B {\bf 207} (1988) 441.

\bibitem{Kounnas:1988ye}
C.~Kounnas and M.~Porrati,
Nucl.\ Phys.\ B {\bf 310} (1988) 355;
S.~Ferrara, C.~Kounnas, M.~Porrati and F.~Zwirner,
Nucl.\ Phys.\ B {\bf 318} (1989) 75.

\bibitem{Hull:1994ys}
C.~M.~Hull and P.~K.~Townsend,
Nucl.\ Phys.\ B {\bf 438} (1995) 109 [arXiv:hep-th/9410167];
E.~Witten,
Nucl.\ Phys.\ B {\bf 443} (1995) 85 [arXiv:hep-th/9503124].

\bibitem{Arkani-Hamed:1998rs}
N.~Arkani-Hamed, S.~Dimopoulos and G.~R.~Dvali,
Phys.\ Lett.\ B {\bf 429} (1998) 263 [arXiv:hep-ph/9803315];
I.~Antoniadis, N.~Arkani-Hamed, S.~Dimopoulos and G.~R.~Dvali,
Phys.\ Lett.\ B {\bf 436} (1998) 257 [arXiv:hep-ph/9804398];
D.~Cremades, L.~E.~Ibanez and F.~Marchesano,
Nucl.\ Phys.\ B {\bf 643} (2002) 93
[arXiv:hep-th/0205074].

\bibitem{KachruJR}
S.~Kachru and A.K.~Kashani-Poor,
JHEP {\bf 0503}, 066 (2005) [arXiv:hep-th/0411279];
J.P.~Derendinger, C.~Kounnas, P.M.~Petropoulos and F.~Zwirner,
arXiv:hep-th/0503229;
G.~Villadoro and F.~Zwirner,
arXiv:hep-th/0503169;
O.~DeWolfe, A.~Giryavets, S.~Kachru and W.~Taylor,
arXiv:hep-th/0505160;
P.G.~Camara, A.~Font and L.E.~Ibanez,
arXiv:hep-th/0506066;
T.W.~Grimm and J.~Louis,
Nucl.\ Phys.\ B {\bf 718}, 153 (2005) [arXiv:hep-th/0412277];
C.~Kokorelis,
Nucl.\ Phys.\ B {\bf 677} (2004) 115
[arXiv:hep-th/0207234].

\bibitem{Kallosh}
S.~Kachru, R.~Kallosh, A.~Linde and S.P.~Trivedi,
Phys.\ Rev.\ D {\bf 68}, 046005 (2003) [arXiv:hep-th/0301240].

\bibitem{Antoniadis:2004pp}
I.~Antoniadis and T.~Maillard,
Nucl.\ Phys.\ B {\bf 716} (2005) 3 [arXiv:hep-th/0412008];
  I.~Antoniadis, A.~Kumar and T.~Maillard,
  arXiv:hep-th/0505260.

\bibitem{Bianchi:2005yz}
M.~Bianchi and E.~Trevigne,
JHEP {\bf 0508} (2005) 034 [arXiv:hep-th/0502147];
M.~Bianchi and E.~Trevigne,
arXiv:hep-th/0506080.

\bibitem{Lust:2005dy}
D.~Lust, S.~Reffert, W.~Schulgin and S.~Stieberger,
arXiv:hep-th/0506090.

\bibitem{Dienes:1998vg}
K.~R.~Dienes, E.~Dudas and T.~Gherghetta,
Nucl.\ Phys.\ B {\bf 537} (1999) 47 [arXiv:hep-ph/9806292].

\bibitem{Gondolo:2004sc}
  P.~Gondolo, J.~Edsjo, P.~Ullio, L.~Bergstrom, M.~Schelke and E.~A.~Baltz,
  JCAP {\bf 0407} (2004) 008
  [arXiv:astro-ph/0406204].

\bibitem{Paige:2003mg}
  F.~E.~Paige, S.~D.~Protopopescu, H.~Baer and X.~Tata,
  arXiv:hep-ph/0312045.

\bibitem{msugra} Hall L J, Lykken J, and Weinberg S 1983 {\it Phys.\ Rev.\ D} {\bf 27} 2359\\
N.~Ohta,
  Prog.\ Theor.\ Phys.\  {\bf 70} (1983) 542.

\bibitem{Barbieri:1982nz}
  R.~Barbieri, S.~Ferrara, L.~Maiani, F.~Palumbo and C.~A.~Savoy,
  Phys.\ Lett.\ B {\bf 115} (1982) 212.

\bibitem{derendinger}
J.P.~Derendinger,
''Globally Supersymmetric Theories in Four and Two Dimensions'',
Proceedings of the Hellenic School of Particle Physics, Corfu, Greece, September 1989, 
edited by G. Zoupanos and N. Tracas, (World Scientific, Singapore, 1990), also available at
http://www.unine.ch/phys/hepth/Derend/derend-frame.html

\bibitem{Martin:1997ns}
S.~P.~Martin,
arXiv:hep-ph/9709356.

\bibitem{prisco}
M.~Prisco,
``Extra dimensions and Dark Matter,''
PhD Thesis Universit\`a di Roma  ``Tor Vergata''.

\bibitem{Kobayashi:1998ye}
T.~Kobayashi, J.~Kubo, M.~Mondragon and G.~Zoupanos,
Nucl.\ Phys.\ B {\bf 550} (1999) 99 [arXiv:hep-ph/9812221].

\bibitem{Yamada:1994id}
Y.~Yamada,
Phys.\ Rev.\ D {\bf 50} (1994) 3537 [arXiv:hep-ph/9401241].

\bibitem{Langlois:2002bb}
D.~Langlois,
Prog.\ Theor.\ Phys.\ Suppl.\  {\bf 148} (2003) 181 [arXiv:hep-th/0209261].

\bibitem{Nihei:2004xv}
T.~Nihei, N.~Okada and O.~Seto,
Phys.\ Rev.\ D {\bf 71} (2005) 063535 [arXiv:hep-ph/0409219].

\bibitem{Okada:2004nc}
N.~Okada and O.~Seto,
Phys.\ Rev.\ D {\bf 70} (2004) 083531 [arXiv:hep-ph/0407092].

\bibitem{Falck:1985aa}
N.~K.~Falck,
Z.\ Phys.\ C {\bf 30} (1986) 247.
\bibitem{Maartens:2003tw}
  R.~Maartens,
  Living Rev.\ Rel.\  {\bf 7} (2004) 7
  [arXiv:gr-qc/0312059].

\bibitem{Kiritsis:2003mc}
  E.~Kiritsis,
  Fortsch.\ Phys.\  {\bf 52} (2004) 200
  [Phys.\ Rept.\  {\bf 421} (2005) 105]
  [arXiv:hep-th/0310001].

\bibitem{Quevedo:2002xw}
  F.~Quevedo,
  Class.\ Quant.\ Grav.\  {\bf 19} (2002) 5721
  [arXiv:hep-th/0210292].

\bibitem{Kofinas:2004ae}
  G.~Kofinas,
  Class.\ Quant.\ Grav.\  {\bf 22} (2005) L47
  [arXiv:hep-th/0412299].

\bibitem{Kofinas:2005py}
  G.~Kofinas,
  arXiv:hep-th/0506035.

\bibitem{Multamaki:2003zc}
  T.~Multamaki and I.~Vilja,
  Phys.\ Lett.\ B {\bf 559} (2003) 1
  [arXiv:hep-th/0301168].

\bibitem{Cuadros-Melgar:2005ex}
  B.~Cuadros-Melgar and E.~Papantonopoulos,
  Phys.\ Rev.\ D {\bf 72} (2005) 064008
  [arXiv:hep-th/0502169].

\bibitem{Apostolopoulos:2004ic}
  P.~S.~Apostolopoulos and N.~Tetradis,
  Phys.\ Rev.\ D {\bf 71} (2005) 043506
  [arXiv:hep-th/0412246];
  P.~S.~Apostolopoulos and N.~Tetradis,
  Phys.\ Lett.\ B {\bf 633} (2006) 409
  [arXiv:hep-th/0509182].

\bibitem{Kiritsis1}
E.~Kiritsis told us he performed unpublished computations that support this assumption.

\bibitem{Langlois:2003yx}
  D.~Langlois,
  arXiv:astro-ph/0301021.

\bibitem{Spergel:2003cb}
  D.~N.~Spergel {\it et al.}  [WMAP Collaboration],
  Astrophys.\ J.\ Suppl.\  {\bf 148} (2003) 175
  [arXiv:astro-ph/0302209].

\bibitem{Tegmark:2003ud}
  M.~Tegmark {\it et al.}  [SDSS Collaboration],
  Phys.\ Rev.\ D {\bf 69} (2004) 103501
  [arXiv:astro-ph/0310723].

\end{thebibliography}
\end{document}